\documentclass[aps,prx,amsmath,amssymb,amsfonts,superscriptaddress,notitlepage,reprint,longbibliography,floatfix
]{revtex4-2}

\usepackage{graphicx}
\usepackage{dcolumn}
\usepackage{bm}
\usepackage{amssymb}
\usepackage{amsmath}
\usepackage[svgnames]{xcolor}
\usepackage[colorlinks=true,
            linkcolor=red,
            urlcolor=blue,
            citecolor=blue]{hyperref}
\usepackage[english]{babel}
\usepackage{braket}
\usepackage{microtype}
\usepackage{bbm}
\usepackage{tikz}
\usepackage{siunitx}
\usepackage{physics}
\usepackage{csquotes}
\usepackage{multirow}
\usepackage{interval}
\usepackage{nicefrac}
\usepackage{caption}
\usepackage{subcaption}
\usepackage{ragged2e}

\usepackage{float}
\makeatletter
\let\newfloat\newfloat@ltx
\makeatother
\usepackage{algorithm}
\usepackage{algpseudocode}
\usepackage[capitalise]{cleveref}

\graphicspath{{./images/}}

\begin{document}

\makeatother

\author{Louis Pallegoix$^1$, Jaime Travesedo$^1$, Alexandre S. May$^{1,2}$, Léo Balembois$^1$, Denis Vion$^1$, Patrice Bertet*$^1$ and Emmanuel Flurin*}

\affiliation{Université Paris-Saclay, CEA, CNRS, SPEC, 91191 Gif-sur-Yvette Cedex, France\\$^2$Alice$\&$Bob, 53 boulevard du Général Martial Valin, 75015 Paris}

\date{\today}

\email{emmanuel.flurin@cea.fr}
\email{patrice.bertet@cea.fr}

\title{Enhancing the sensitivity of single-microwave-photon detection with bandwidth tunability.}

\begin{abstract}
We report on the characteristics of a microwave photon counter device based on a superconducting transmon qubit. Its design is similar to~\cite{balembois_cyclically2024}, with an additional bandwidth tuning circuit that permits tuning the detection bandwidth over an order of magnitude to optimize the device efficiency and noise. Owing to this new feature and to improvements in device fabrication, a power sensitivity of $3 \cdot 10^{-23}\ \mathrm{W}/\sqrt{\mathrm{Hz}}$ is reached. We confirm the high performance of the device by measuring single spin microwave fluorescence. 
\end{abstract}

\maketitle

\section{Introduction}

Single-photon detection has long been a fundamental tool in quantum technologies, with well-established applications in the optical domain, ranging from fluorescence microscopy~\cite{orrit_single_1990,klar_fluorescence_2000,betzig_imaging_2006,bruschini_single-photon_2019} to quantum communication~\cite{gisin_quantum_2002} and boson sampling~\cite{banchi_molecular_2020} ; however, extending single-photon detection into the microwave regime presents significant challenges. The energy of microwave photons—of the order of µeV—is several orders of magnitude lower than that of optical photons, requiring detectors to operate at cryogenic temperatures to suppress thermal noise and reduce the background thermal photon population.

Despite these challenges, single-microwave-photon detectors (SMPDs) have emerged as critical tools for advancing quantum sensing and information processing. Their ability to detect individual microwave photons has opened possibilities for a wide range of applications, including the detection of weak incoherent microwave emitters, such as electron spins in solids~\cite{albertinale_detecting_2021,wang_single-electron_2023,billaud_fluorescence-detection_2023}, and even searches for hypothetical dark-matter particles~\cite{lamoreaux_analysis_2013,dixit_searching_2021}. SMPDs also hold potential for primary thermometry at low temperatures~\cite{scigliuzzo_primary_2020} and for use in quantum illumination protocols~\cite{assouly_quantum_2023}. Moreover, these detectors can play a key role in quantum information processing~\cite{raussendorf_measurement-based_2003,briegel_measurement-based_2009,bartolucci_fusion-based_2021}, enabling heralded entanglement between superconducting qubits~\cite{narla_robust_2016}, improving qubit readout schemes~\cite{opremcak_measurement_2018}, and facilitating the generation of robust quantum states~\cite{besse_parity_2020}.

Various SMPD designs have been proposed and implemented, with most relying on superconducting qubits~\cite{lescanne_irreversible_2020,albertinale_detecting_2021,narla_robust_2016,romero_microwave_2009,helmer_quantum_2009,sathyamoorthy_quantum_2014,kyriienko_continuous-wave_2016,sathyamoorthy_detecting_2016,gu_microwave_2017,royer_itinerant_2018,chen_josephson-junctions-based_2011,koshino_impedance-matched_2013,inomata_single_2016,besse_single-shot_2018,kono_quantum_2018}, quantum dots~\cite{wong_double-quantum-dot_2017,Ghirri_microwave_2020} or bolometric detectors~\cite{lee_graphene-based_2020}. These devices have demonstrated promising results, achieving sensitivities ranging from $10^{-21}$ to $10^{-22} \, \mathrm{W/\sqrt{Hz}}$ at microwave frequencies ; however, there remain limitations in terms of bandwidth, detection efficiency, and dark count rates that must be addressed to further enhance their performance. 

In this work, we present an enhanced version of the SMPD introduced in Ref.~\cite{balembois_cyclically2024}, which is based on a superconducting qubit and a four-wave mixing (4WM) process. The design integrates a frequency-tunable Purcell filter, providing control over the detection bandwidth. Since the thermal noise power is proportional to the detection bandwidth, this tunability significantly reduces thermal noise and decreases the dark count rate to $30\ \mathrm{s}^{-1}$, improving the overall detector performance. Additionally, advancements in the fabrication process have led to an extended qubit relaxation time ($T_1$), which in turn enhances the detection efficiency to $0.8$. As a result of these improvements, the detector achieves a power sensitivity of $3 \cdot 10^{-23} \, \mathrm{W/\sqrt{Hz}}$, representing a significant advancement over previous designs. We confirm the improved sensitivity of the SMPD by detecting the microwave fluorescence from a single spin. 

\section{SMPD principle of operation}

\subsection{Device design}

The SMPD device (see Figs.~\ref{fig:device_overview}a and b) consists of a transmon qubit (mode $q$, frequency $\omega_q$), capacitively coupled to two harmonic oscillator modes, called the buffer (mode $b$, frequency $\omega_b$), and the waste (mode $w$, frequency $\omega_w$). The buffer is connected to the input line of the device through another resonator at frequency $\omega_{pb}$. The latter serves as Purcell filter to decouple the qubit from the input line, while also setting the effective buffer energy coupling rate $\kappa_{\mathrm{b,c}}$. The total energy-loss rate of the buffer resonator is $\kappa_b = \kappa_{\mathrm{b,c}}+\kappa_{\mathrm{b,i}}$, where $\kappa_{\mathrm{b,i}}$ represents the internal losses of the buffer. The waste is connected to an output line through another Purcell filter, at frequency $\omega_{pw}$, with a rate $\kappa_w$. Other relevant parameters are the qubit energy-relaxation time $T_1$ and equilibrium excited-state population, $p_{\mathrm{th},q}$, which varies with the qubit temperature $T$, in the $T \ll \hbar\omega/k_B$ limit, as $p_{\mathrm{th},q} = \frac{1}{e^{\frac{\hbar \omega_q}{k_{B}T}}-1}$. In this formula, $T$ is not necessarily the cryostat base temperature, as various heating effects (due to infrared radiation in particular) may lead to a higher effective qubit temperature.

For optimal operation as an energy detector, both the frequency and bandwidth of the device need to be matched to the source of interest, be it a spin~\cite{balembois_cyclically2024, wang_single-electron_2023} or a dark-matter candidate~\cite{braggio_quantum-enhanced_2024}. Frequency tuning is achieved by inserting a superconducting quantum interference device (SQUID) into the buffer mode, resulting in a buffer frequency $\omega_{b}(\phi_b)$ that depends on the flux threading the buffer SQUID loop, $\phi_b$. An asymmetric SQUID is purposely chosen to minimize sensitivity to flux noise. Bandwidth tuning is achieved by inserting a second SQUID into the Purcell filter mode, resulting in a buffer coupling rate $\kappa_{\mathrm{b,c}}(\phi_{\mathrm{pb}})$ that depends on the flux threading the Purcell SQUID loop, $\phi_{\mathrm{pb}}$. This SQUID is symmetric to maximize the bandwidth-tuning range.

\subsection{Four-wave mixing}

\begin{figure}[htpb!]
\includegraphics[scale=0.4, trim = 5.3cm 3.5cm 4.6cm 3.5cm, clip]{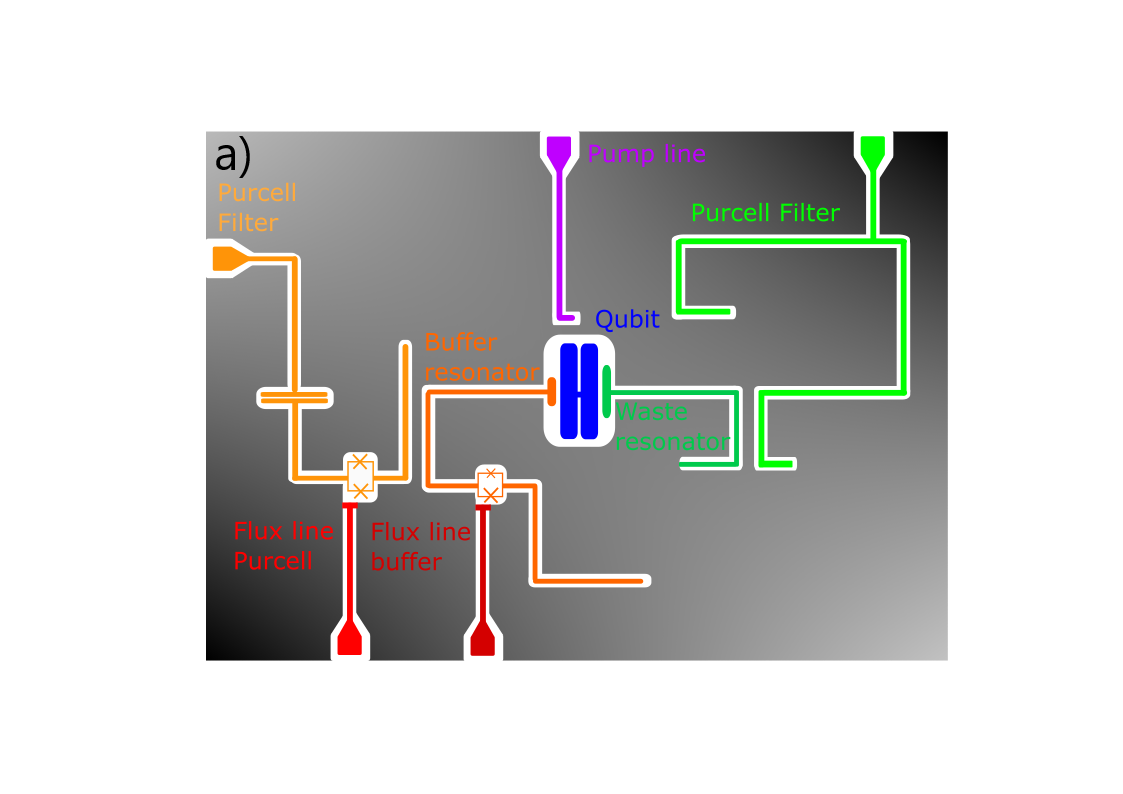}
\includegraphics[scale=0.28, trim = 0cm 3.5cm 0cm 4cm, clip]{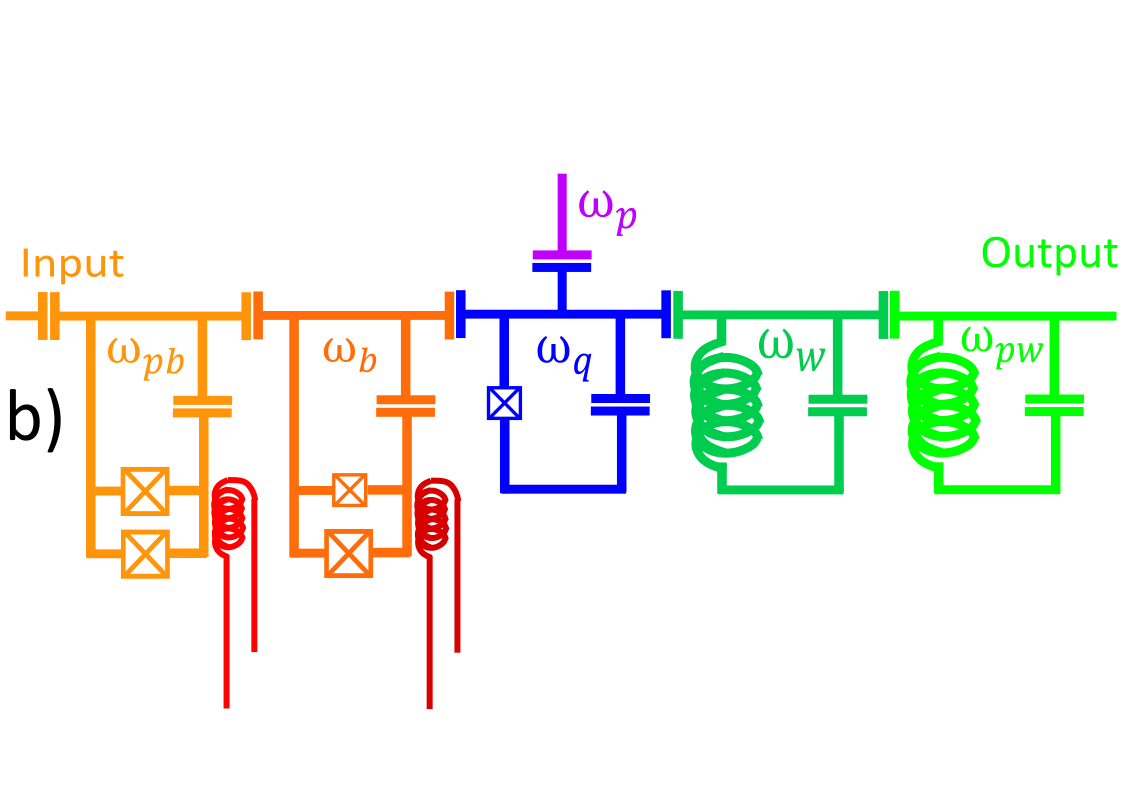}
\includegraphics[scale=0.28, trim = 0cm 2cm 0cm 0cm, clip]{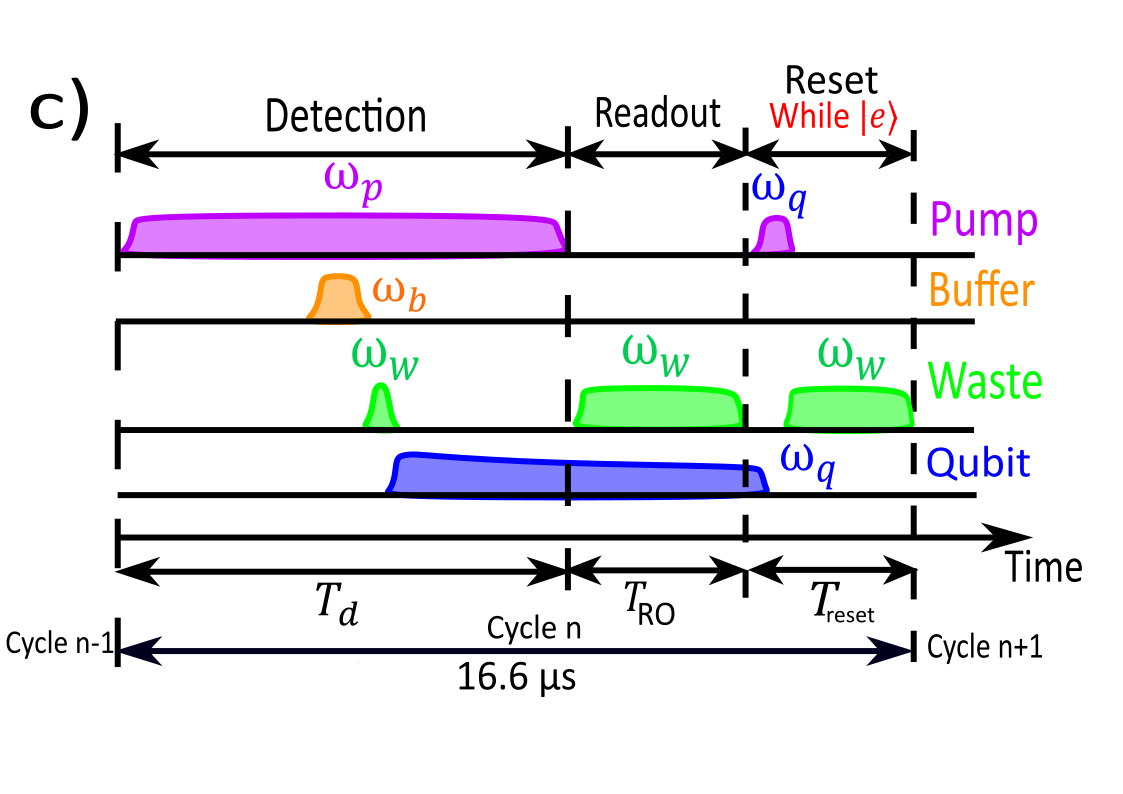}
\captionsetup{justification=Justified}
\caption{Overview of the device. a) Global layout of the SMPD. b) Lumped-element scheme of the SMPD. The colors of the different resonators are in accordance with their colors on the chip scheme. Note : The buffer SQUID is asymmetrical. c) Operation cycle of the SMPD. During the detection step (duration $T_d$), the pump is switched on; a photon arriving in the buffer mode is converted into a photon in the waste, which quickly decays, and a qubit excitation, which decays in a characteristic time $T_1 \gg T_d$. Dispersive readout is performed through the waste resonator in time $T_\mathrm{RO}$. If the qubit is found in $\ket{e}$, a $\pi$ pulse is applied, followed by qubit readout, and reapplication of the reset if it is found in $\ket{e}$. This conditional reset step lasts $T_\mathrm{reset}$. The whole cycle lasts $16.6$ µs (assuming one reset cycle). }
\label{fig:device_overview}
\end{figure}

The working principle of photon detection relies on a 4WM process that maps the presence or absence of a photon onto the state of the transmon qubit. The nonlinear element is provided by the qubit Josephson junction. The 4WM is activated by a microwave signal (called the pump, frequency $\omega_p$) of amplitude $\xi_0$, applied to the qubit via a pump line connected capacitively to the qubit (see Fig.~\ref{fig:device_overview}). When the pump frequency approximately satisfies the relation $\omega_b + \omega_p = \omega_q+\omega_w-\chi_w$, the dominant process is described by the Hamiltonian

\begin{equation}
    H_{\mathrm{4WM}} = \sqrt{\chi_b \chi_w} (\xi_0 b q^\dagger w^\dagger + \xi_0^* b^\dagger q w), 
\end{equation}

where $\chi_b$ (resp. $\chi_w$) is the qubit dispersive coupling to the buffer mode (resp. the waste mode)~\cite{lescanne_irreversible_2020}. This Hamiltonian reversibly converts a buffer and pump photon into a qubit excitation and waste photon. Dissipation of the waste photon into the output line ensures that the process is irreversible ; nonetheless, the qubit remains excited long enough to be readout subsequently, and it can thus be used as a flag that frequency conversion occurred. 

An approximate description is obtained by eliminating the qubit degree of freedom, yielding an effective Hamiltonian $H_{\mathrm{eff}} = g_{\mathrm{4WM}} b w^\dagger + h.c.$ with $g_{\mathrm{4WM}} = \sqrt{\chi_b \chi_w} \xi_0$. This describes the parametric coupling between the buffer and waste modes, similar to the Josephson parametric converter for instance. Assuming that $\kappa_{\mathrm{b,i}} \ll \kappa_{\mathrm{b,c}}$, the efficiency of frequency conversion under this process is $\eta_{\mathrm{4WM}} = 4 C / (1+C)^2$, where $C = \frac{4 \chi_b \chi_w |\xi_0|^2}{\kappa_w \kappa_b}$ is the system cooperativity. Unit efficiency is reached when $C=1$. Physically, this occurs when the frequency-conversion rate $4 |g_{\mathrm{4WM}}|^2/\kappa_w $ is equal to the input resonator coupling rate, $\kappa_b$. This condition can always be achieved by adjusting the pump frequency and amplitude (see Appendix A for more details).

\subsection{Cyclic operation}

The SMPD is operated cyclically. A cycle consists of three successive steps: detection, readout, and reset (see Fig. \ref{fig:device_overview}c. During the detection step, which lasts $T_d$, the pump drive is simply switched on, with suitable amplitude and frequency for maximum frequency conversion. In the next step, dispersive qubit readout is achieved via the waste resonator by measuring the reflected field quadratures of a microwave tone in a time $T_\mathrm{RO}$. We define the readout fidelity $\mathcal{F}_{\mathrm{RO}}^{\ket{e}} < 1$ as the probability of measuring the qubit in the excited state after a $\pi$ pulse. 

In the third step, the qubit is reset into its ground state by applying a $\pi$ pulse conditioned to the outcome of the previous readout being the excited state. This efficiently resets the qubit into its ground state, with a probability that can be even lower than the thermal-equilibrium qubit excited-state probability~\cite{balembois_cyclically2024}. Note that the duration of the reset step $T_\mathrm{reset}$ varies from cycle to cycle due to its conditional nature. The detector duty cycle $\eta_\mathrm{cycle}$ is defined as the average value of $T_d / (T_d + T_\mathrm{RO} + T_\mathrm{reset})$.

\section{Figures of merit}

In this section, we describe in some detail the processes that limit the performance of the device as an energy detector. This performance is captured by the power sensitivity $S$, defined as the power that can be detected with unit signal-to-noise ratio in $1$\,s integration time~\cite{balembois_cyclically2024}:

\begin{equation}
S= \hbar\omega_{b}\frac{\sqrt{\alpha}}{\eta_\mathrm{SMPD}}.
\label{eq:Sensitivity}
\end{equation}

In this formula, $\eta_\mathrm{SMPD}$ is the overall detection efficiency of the SMPD and $\alpha$ is the average number of counts per second when no signal is sent to the device. The latter can be separated in two contributions $\alpha = \alpha_\mathrm{err} +  \alpha_{\mathrm{th}}$, $\alpha_\mathrm{err}$ being the false-positive events and $\alpha_{\mathrm{th}}$ being the counts due the detection of thermal photons in the input line. In what follows, we address each of these three quantities, with the goal of maximizing $\eta_\mathrm{SMPD}$ while minimizing $\alpha_\mathrm{err}$ and $\alpha_{\mathrm{th}}$.

\subsection{Detection efficiency}

A photon can be missed if :

\begin{itemize}
\item It is spectrally filtered by the detector
\item The 4WM efficiency is not $1$
\item The qubit relaxes back in its ground state during the detection step
\item Qubit excited state readout fails
\item The photon hits the detector during the readout or reset steps
\end{itemize}

Therefore, the probability that a photon in the input line at frequency $\omega$ gives rise to a count is

\begin{equation}
\eta_\mathrm{SMPD}(\omega) = \eta_\omega \eta_{\mathrm{4WM}} \eta_q \mathcal{F}_{\mathrm{RO}}^{\ket{e}} \eta_{\mathrm{cycle}},
\label{eq:Efficiency}
\end{equation}

\noindent where  the  detector frequency response function is approximated by a Lorentzian of width $\kappa_d$ (the detector linewidth) such that $\eta_\omega = \frac{1}{1+\left[ \frac{2 (\omega- \omega_b)}{\kappa_d} \right]^2}$ .
The qubit efficiency $\eta_q = \frac{T_1}{T_d} (1 - \mathrm{e}^{-\frac{T_d}{T_1}})$ can be approximated by $1 - \frac{T_d}{2T_1}$ in the limit $T_d \ll T_1$~\cite{balembois_cyclically2024}. To maximize $\eta_\mathrm{SMPD}$, $T_d$ should be optimized to maximize the duty cycle while keeping the cycle time well below $T_{1}$ to avoid relaxation of the qubit before readout.

\subsection{Dark count rate}

The dark count rate $\alpha_\mathrm{err}$ is the number of counts detected per unit of time, even though no photon was present in the input line. It has two contributions, $\alpha_\mathrm{err} = \alpha_q + \alpha_p$. First, the qubit can be found in the excited state despite the active reset procedure. This is due to the relaxation toward the thermal-equilibrium value $p_{q,th}$, which takes place during the detection window with a characteristic time $T_1$~\cite{balembois_cyclically2024}, giving rise to $\alpha_q = p_{th,q} (T_d / T_1) (1/(T_d + T_\mathrm{RO} + T_\mathrm{reset}))$ (assuming that the active reset works perfectly, otherwise an offset must be added to that expression). The other contribution $\alpha_p$ is due to heating brought by the pump. Note that the readout errors causing the ground state to be erroneously detected as the excited state are negligible because we purposely choose a detection threshold that minimizes this effect. 

\subsection{Thermal counts}

Thermalizing the microwave field at millikelvin temperatures is well known to be a difficult task because of vanishing thermal conductivity and the large attenuation needed in the input lines. This results in an effective field temperature that is often significantly larger than the cryostat base temperature, and in our case, in the presence of a finite flux of thermal photons at the SMPD input, causing a thermal count rate of thermal origin

\begin{equation}
\alpha_{\mathrm{th}} = \frac{1}{2\pi} \int_{-\infty}^{\infty}  \Bar{n}_{\mathrm{th,b}} \eta_\mathrm{SMPD}(\omega) d\omega,
\label{eq:th_integral}
\end{equation}

\noindent where $\Bar{n}_{\mathrm{th,b}} = \frac{1}{e^{\frac{\hbar \omega_b}{k_{B}T}}-1}$ is the number of photons per mode assuming that the field is at temperature $T$. Because $\kappa_d \ll \omega_b$, we get

\begin{equation}
\alpha_{\mathrm{th}} = \frac{\Bar{n}_{\mathrm{th,b}}  \kappa_{d}\eta_\mathrm{SMPD}(\omega_b)}{4}.
\label{eq:th_countrate}
\end{equation}

Due to the tunable bandwidth of our device, we are able here to test the predicted proportionality between noise and detector bandwidth, as will be seen below. 

\section{Measurements}

\subsection{Device characterization}

The SMPD resonators are measured with a vector network analyser (VNA). Each resonance is fitted to determine the resonator frequency and linewidth. The results of these fits are reported in Tab. \ref{tab:f&Q}.

\begin{table}
    \centering
    \begin{tabular}{| c | c | c |}
        \hline
        Resonator & Frequency & Linewidth \\ \hline
        Buffer Purcell & $\omega_{pb}/2\pi \in [7.26 ; 7.78]$  GHz & 20 MHz \\ \hline
        Buffer & $\omega_{b}/2\pi \in [7.70 ; 7.76]$ GHz & See fig. \ref{fig:bdw_tuning} \\ \hline
        Qubit & $\omega_{q}/2\pi = 6.533$ GHz & 16 kHz \\ \hline
        Waste & $\omega_{w}/2\pi = 8.475$ GHz & 1.75 MHz \\ \hline
        Waste Purcell & $\omega_{pw}/2\pi \approx 8.39$ GHz & 400 MHz \\ \hline
    \end{tabular}
    \captionsetup{justification=Justified}
    \caption{Measured resonator frequencies and linewidths. The Purcell filter of the buffer and the buffer itself are tunable over the indicated interval. The buffer linewidth depends on the the Purcell frequency, as shown in fig. \ref{fig:bdw_tuning}.}
    \label{tab:f&Q}
\end{table}

\begin{table}
    \centering
    \begin{tabular}{| c | c | c |}
        \hline
        Resonator & Cross-Kerr to qubit \\ \hline
        Buffer & $\chi_b/2\pi  = 3.5$ MHz \\ \hline
        Waste & $\chi_w/2\pi = 16$ MHz \\ \hline
    \end{tabular}
    \captionsetup{justification=Justified}
    \caption{Measured dispersive shifts induced by the qubit to its neighboring resonators.}
    \label{tab:chis_table}
\end{table}

\begin{figure}[htpb!]
\includegraphics[scale = 0.41]{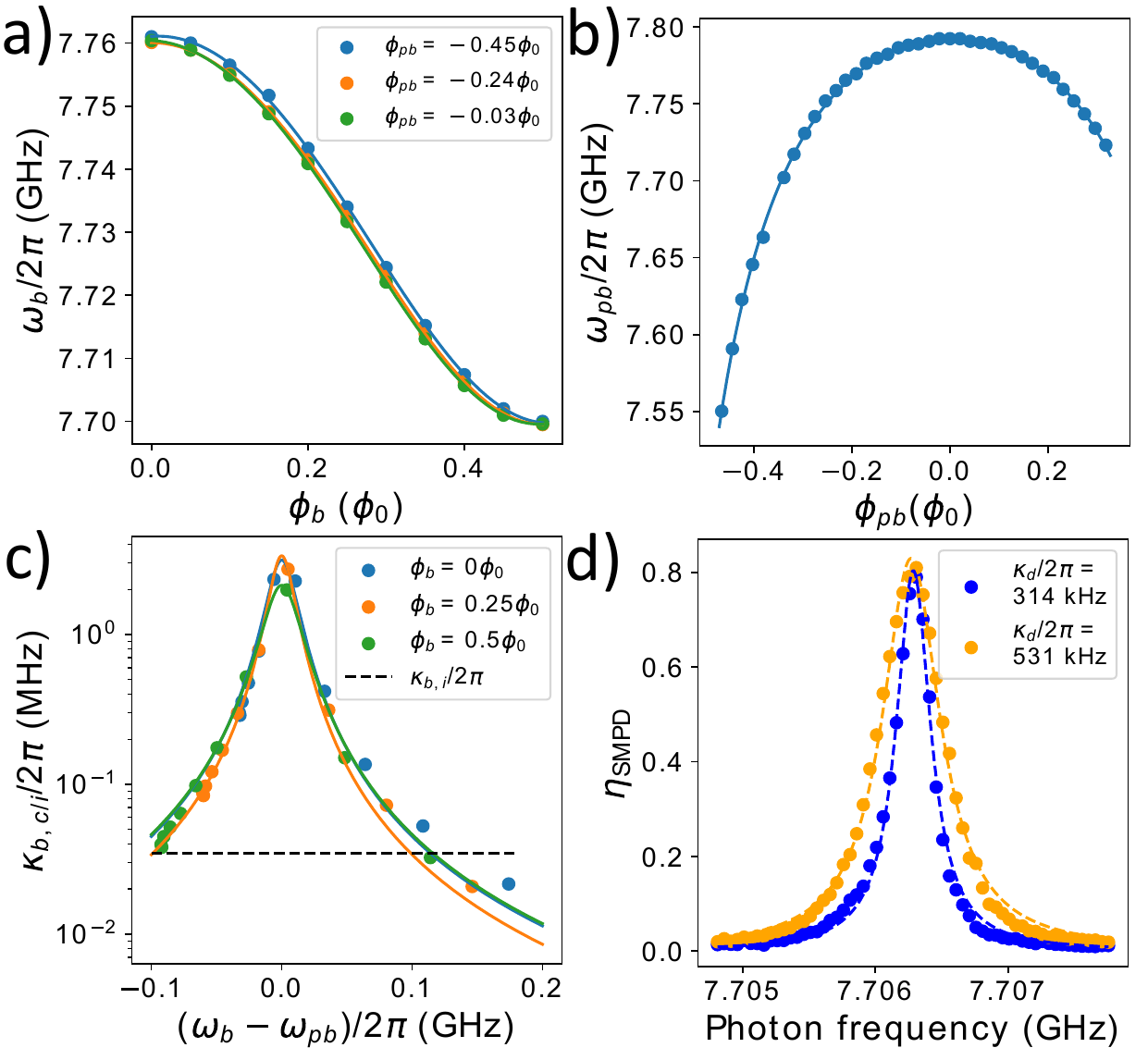}
\captionsetup{justification=Justified}
\caption{Tunability curves. a) Solid circles are the measured frequency of the buffer resonator, $\omega_b$, as a function of the flux bias applied to its SQUID, $\Phi_b$, for several values of $\Phi_{pb}$ (blue is $-0.45 \Phi_0$, orange is $-0.24 \Phi_0$, green is $-0.03 \Phi_0$). The solid lines are sinusoidal fits, yielding the asymmetry between the buffer SQUID junctions. b) Full blue circles are the measured frequency of the Purcell filter, $\omega_{pb}$, as a function of the flux bias sent to its SQUID $\Phi_{pb}$. The solid line is a fit assuming a symmetric SQUID. c) Full circles show the measured coupling rate of the buffer resonator to the $50 \Omega$ line, as a function of its detuning from the Purcell filter, $\omega_b - \omega_{pb}$, for $3$ different values of $\Phi_b$ : blue for $\Phi_b = 0$, orange for $\Phi_b = 0.25 \Phi_0$, green for $\Phi_b = 0.5 \Phi_0$. The dashed line shows the internal losses of the buffer, $\kappa_{b,i}$. The solid lines are lorentzian fits to data of the same color. We can see that $\kappa_b/2\pi$ rise above 3 MHz at zero detuning and fall below $\kappa_{b,i}$ when the detuning exceeds 100 MHz in absolute value. d) Full circles are the measured SMPD efficiency, $\eta_\mathrm{SMPD}$, as a function of signal frequency $\omega$, for $\kappa_b/2\pi = 170$\,kHz (blue) and $280$\,kHz (orange). Dashed lines are Lorentzian fits yielding the SMPD bandwidth $\kappa_d /2\pi$, equal to $314$\,kHz and $531$\,kHz, respectively.
}
\label{fig:bdw_tuning}
\end{figure}

Figure ~\ref{fig:bdw_tuning}a shows $\omega_{b}(\phi_b)$ for several values of $\phi_{pb}$. The buffer tuning range is narrow (60 MHz) and the curve is well fitted by a sinusoid~\cite{blais_cqed_2021}, from which we deduce a ratio of 15 between the Josephson-junction critical currents of the asymmetric SQUID. Little crosstalk between the $\phi_b$ and $\phi_{pb}$ is observed, as evidenced by the weak dependence of $\omega_{b}/2\pi$ on $\phi_{pb}$.\newline

From the fits of the buffer resonance curves, we extract $\kappa_{b,i} = 2.2 \cdot 10^5 \mathrm{s}^{-1}$, and $\kappa_{b,c}$, the internal loss and coupling rates, respectively. Fig.~\ref{fig:bdw_tuning}c shows $\kappa_{b,c}/2\pi$ as a function of the detuning from the Purcell, $\omega_b-\omega_{pb}$. A large tuning range is observed, ranging from $\sim 10$\,kHz up to $\sim 3$\,MHz. The tuning curve is identical for all values of buffer-resonator frequency (see Fig.~\ref{fig:bdw_tuning}c). The buffer is overcoupled over a large parameter range (corresponding to $\kappa_{b,c} \gg \kappa_{b,i}$), which is desirable for SMPD operation, as buffer internal losses further reduce $\eta_{\mathrm{4WM}}$~\cite{balembois_cyclically2024}. 

Qubit readout is performed at the waste-resonator frequency when the qubit is in the excited state, $\omega_w - \chi_w$, to minimize the chances of exciting it out of the ground state and causing unwanted dark counts, with a readout duration $T_\mathrm{RO} = 0.8$ µs. Readout histograms are shown in Fig.~\ref{fig:qubit_parameters}a (in appendix), yielding the readout fidelity $\mathcal{F}_{\mathrm{RO}}^{\ket{e}} = 0.87$ and equilibrium excited-state population $p_{\mathrm{th},q} = 8.5 \cdot 10^{-4}$. Note that the readout fidelity is significantly lower than those found in recent experiments, which reach values close to 0.99~\cite{walter_realizing_2017}. This is because the waste-resonator parameters ($\chi_w$ and $\kappa_w$) are optimized for the 4WM process and not for qubit readout, which would require a larger value of $\kappa_w$, reaching $\sim \chi_w$~\cite{walter_realizing_2017} (see tab. \ref{tab:f&Q} and \ref{tab:chis_table}). The qubit lifetime is $T_1 = 70$ µs (see Fig.~\ref{fig:qubit_parameters}b), and the coherence time measured under a Ramsey sequence is $T_2^* = 20$ µs. The dispersive shifts $\chi_b$ and $\chi_w$ are reported in Table \ref{tab:chis_table}. Active reset is performed as explained in Section IIC~\cite{balembois_cyclically2024}.

\subsection{Four-wave-mixing tuning}

The 4WM condition is found by sending a pump tone at frequency $\omega_p$ together with a signal at frequency $\omega$ into the SMPD input, followed by qubit readout. Sweeping $\omega_p$ and $\omega$, the 4WM condition is identified by an enhanced qubit excited state when the signal is present, which vanishes when it is switched off (see Fig.~\ref{fig:4WM}) ; the pump power is then tuned to reach unit cooperativity (see Appendix and Fig.~\ref{fig:4WM}).

\subsection{Cyclic operation : Detector bandwidth and dark counts}

We now turn to the operation of the SMPD in cyclic mode. The detection step duration is set to $T_d = 15$ µs. When no photon is sent to the device, the qubit is most often in the ground state at the end of the sequence, implying that no reset is needed, and that the average cycle duration is very close to $T_d + T_\mathrm{RO} = 15.8$ µs, yielding a duty cycle $\eta_\mathrm{cycle} =0.95$. From Eq.~\ref{eq:Efficiency}, we expect $\eta_\mathrm{SMPD}$ to reach $0.74$.

The experimental $\eta_\mathrm{SMPD}$ is measured by counting the number of clicks detected over 1 second integration time, at the same time that a calibrated signal is sent to the SMPD input. The signal power is calibrated using the qubit as a detector (see \cite{balembois_cyclically2024} and Appendix B). The result is plotted as a function of input frequency $\omega$ in Fig.~\ref{fig:bdw_tuning}d. The maximum efficiency reaches $\eta_\mathrm{SMPD} = 0.8$. This is even slightly higher than the predicted value, which implies in particular that the 4WM efficiency $\eta_{\mathrm{4WM}}$ is very close to $1$. Fitting $\eta_\mathrm{SMPD} (\omega) $ with a lorentzian, we obtain the detector bandwidth $\kappa_d$. Two curves are shown, obtained for two different values of $\kappa_b$. The maximum efficiency is unchanged, but the detector bandwidth is changed according to $\kappa_b$. This confirms that the SMPD bandwidth can be adjusted by tuning $\kappa_b$ via the Purcell filter frequency. We measure $\kappa_d/2\pi = 315$\,kHz for $\kappa_b/2\pi = 170$\,kHz, and $\kappa_d/2\pi = 530$\,kHz for $\kappa_b/2\pi = 280$\,kHz, thus confirming the approximate relation $\kappa_d \simeq 2 \kappa_b$ (see Appendix A).

Dark counts are measured by recording photon-counting traces in the absence of any input signal (see Tab. \ref{tab:bins}, where the detector bandwidth was $\kappa_d/2\pi = 170$\,kHz). To obtain the various contributions to the dark counts, the traces are systematically measured with the pump off (yielding $\alpha_q$), with the pump on and detuned by $20$\,MHz (yielding $\alpha_q + \alpha_p = \alpha_\mathrm{err}$), and with the pump on and tuned to the 4WM conditions (yielding $\alpha = \alpha_\mathrm{err} + \alpha_{\mathrm{th}}$). At $10$\,mK, we obtain $\alpha_q = 8 \mathrm{s}^{-1}$, $\alpha_p = 2 \mathrm{s}^{-1}$, and $\alpha_{\mathrm{th}} = 21 \mathrm{s}^{-1}$. Combining this with the measured efficiency, we obtain a power sensitivity $S = 3 \cdot 10^{-23} \mathrm{W}/\sqrt{\mathrm{Hz}}$, a factor of 3 better than the previous state of the art~\cite{balembois_cyclically2024}, which corresponds to an order of magnitude gain in measurement time.

\begin{figure}[htpb!]
\includegraphics[scale=0.575]{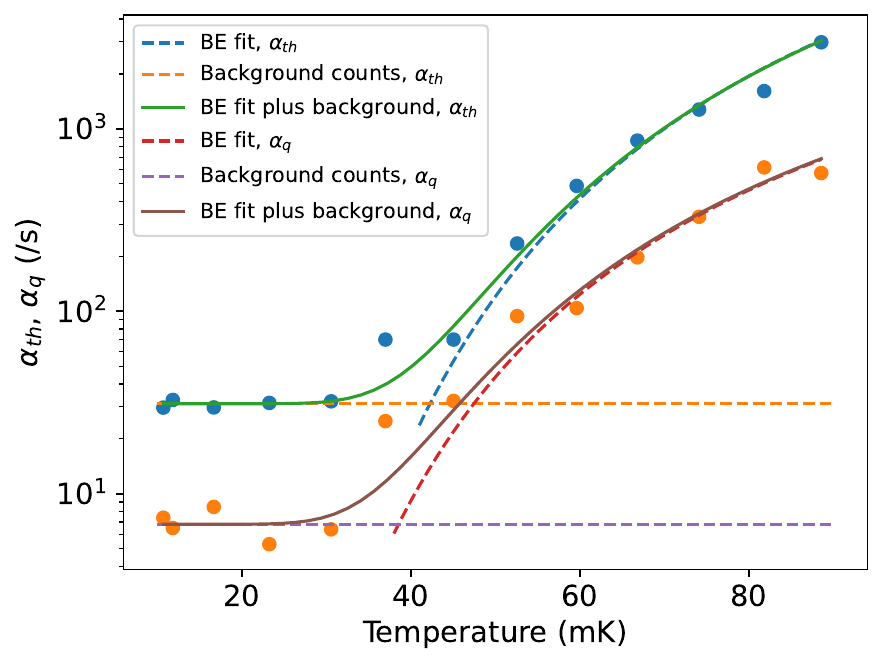}
\captionsetup{justification=Justified}
\caption{Measured $\alpha_{\mathrm{th}}$ (blue full circles) and $\alpha_q$ (orange full circles) as a function of the cryostat's temperature $T$, for a detection bandwidth $\kappa_d/2\pi = 170$\,kHz. The solid green line is a fit of $\alpha_{\mathrm{th}}$ to $\alpha_{\mathrm{th},0} + K_{\mathrm{th}} \bar{n}_{\mathrm{th},b}(T)$, yielding $\alpha_{\mathrm{th},0} = 31 \mathrm{s}^{-1}$ and $K_{\mathrm{th}} = 2 \cdot 10^5 \mathrm{s}^{-1}$. The solid brown line is a fit of $\alpha_q$ to $\alpha_{q,0} + K_q p_{\mathrm{th},q}(T)$, yielding $\alpha_{q,0} = 7 \mathrm{s}^{-1}$ and $K_q = 2.2 \cdot 10^4 \mathrm{s}^{-1}$. The dashed lines are the separated Bose-Einstein and constant components of the respective fits.}
\label{fig:BE}
\end{figure}

Time traces are similarly measured by varying the cryostat temperature $T$. The values of $\alpha_q$ and $\alpha_{\mathrm{th}}$ are shown as a function of $T$ in Fig.~\ref{fig:BE}. The qubit contribution increases with temperature because of the higher excited-state probability at equilibrium; it is well fitted by $\alpha_{q,0} + K_q p_{\mathrm{th},q}(T)$, with $\alpha_{q,0} = 7\, \mathrm{s}^{-1}$ and $K_q = 2.2 \cdot 10^4 \,\mathrm{s}^{-1}$. The value of $K_q$ is slightly larger than the predicted $(T_d / T_1)(1/(T_d + T_\mathrm{RO} + T_\mathrm{reset}))$ (see Section IIIB), which is $1.4 \cdot 10^4 \,\mathrm{s}^{-1}$, possibly because of a reduction of $T_1$ at higher temperature. The thermal contribution also increases with temperature; it is well fitted by $\alpha_{\mathrm{th},0} + K_{\mathrm{th}} \bar{n}_{\mathrm{th},b}$, with $\alpha_{\mathrm{th},0} = 31 \mathrm{s}^{-1}$, and $K_{\mathrm{th}} = 2 \cdot 10^5 \, \mathrm{s}^{-1}$. This value is in quantitative agreement with the expected $\eta_\mathrm{SMPD} \kappa_d / 4 = 2.1 \cdot 10^5 \mathrm{s}^{-1}$, thus providing an independent confirmation of the efficiency estimate reported earlier.

The equilibrium qubit excitation (without active reset) $p_{\mathrm{th},q}$ at $10$\,mK translates into an effective qubit temperature of $42$\,mK. The thermal photon rate $\alpha_{\mathrm{th},0}$ measured at $10$\,mK translates into a mean photon number per mode of $1.5 \cdot 10^{-4}$ and thus to an effective temperature of $40$\,mK. This is higher than the mean photon number expected from our line attenuation, which we estimate to be below $\sim 2 \cdot 10^{-5}$. Our results therefore confirm the well-known difficulty of thermalizing microwave fields and modes to temperatures much lower than $\sim 40$\,mK, possibly due to the difficulty of perfectly thermalizing microwave components. It is interesting to note that SMPD devices are ideally well suited to studying this phenomenon, which represents a major issue for superconducting quantum computers.

\begin{table*}
    \centering
    \begin{tabular}{| c | c | c | c | c |}
        \hline
        Temperature & Pump state & $\kappa_{d}/2\pi$ & One-second click sequence & Click rate \\ \hline
        10 mK & Off & Not relevant & \includegraphics[scale=0.27]{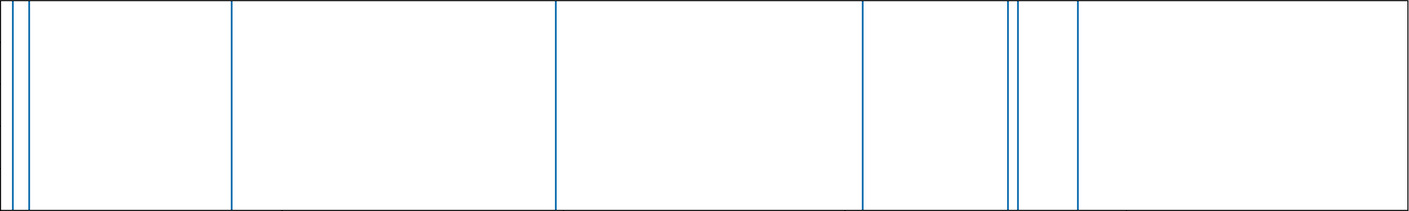} & 8/s \\ \hline
        10 mK & Detuned & Not relevant & \includegraphics[scale=0.27]{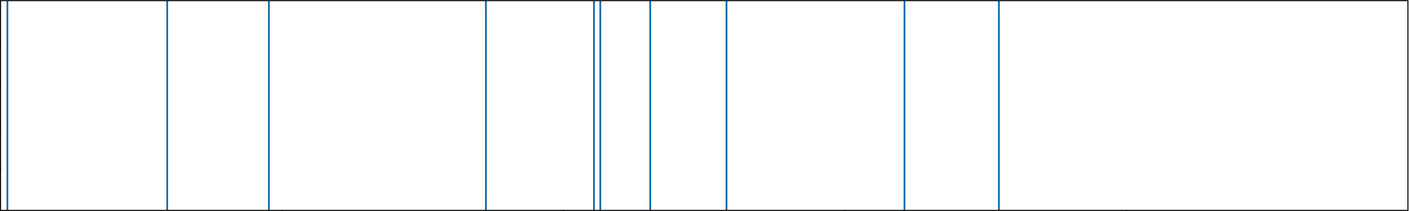} & 10/s \\ \hline
        10 mK & Tuned & 170 kHz & \includegraphics[scale=0.27]{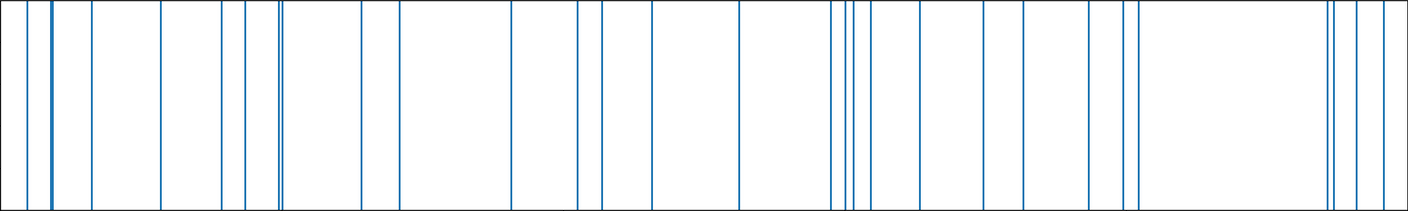} & 31/s \\ \hline
        50 mK & Tuned & 170 kHz & \includegraphics[scale=0.27]{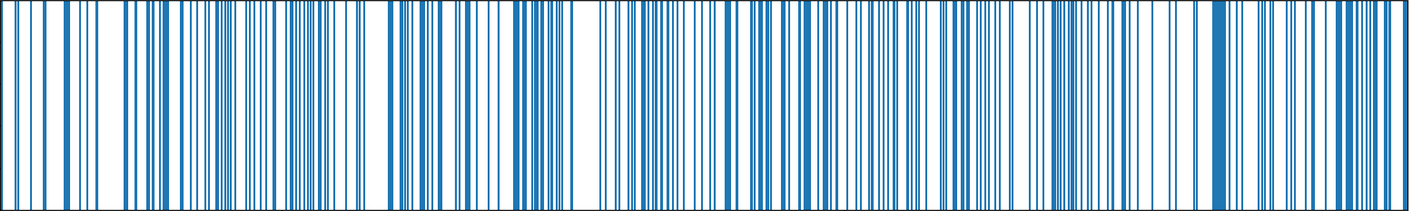} & 344/s \\ \hline
        60 mK & Tuned & 170 kHz & \includegraphics[scale=0.27]{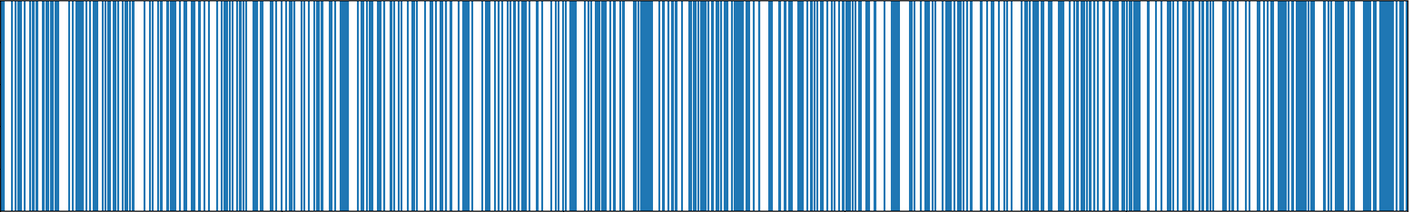} & 621/s \\ \hline
        90 mK & Tuned & 170 kHz & \includegraphics[scale=0.27]{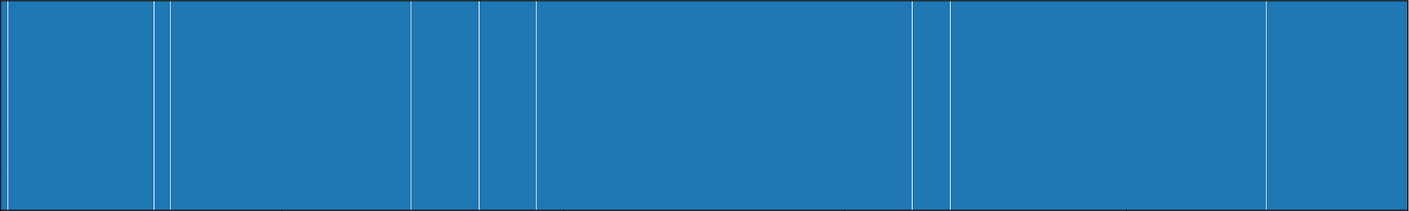} & 3614/s \\ \hline
    \end{tabular}
    \captionsetup{justification=Justified}
    \caption{Photon-counting traces taken under different conditions. First line : the mixing-chamber temperature is 10 mK and the pump is off, yielding $\alpha_q$. Second line : the mixing-chamber temperature is 10 mK and the pump is switched on, but it is detuned by $20$\,MHz from the 4WM condition, yielding $\alpha_\mathrm{err} = \alpha_q + \alpha_p$. Third line : the mixing-chamber temperature is 10 mK and the pump is on and in the 4WM condition, yielding $\alpha_\mathrm{err} + \alpha_{\mathrm{th}}$. Three following lines : the mixing-chamber temperature is increased, and the count rate rises accordingly, as detailed in fig. \ref{fig:BE}.}
    \label{tab:bins}
\end{table*}

\subsection{Dependence on detection bandwidth}

\begin{figure}[htpb!]
\includegraphics[scale=0.855]{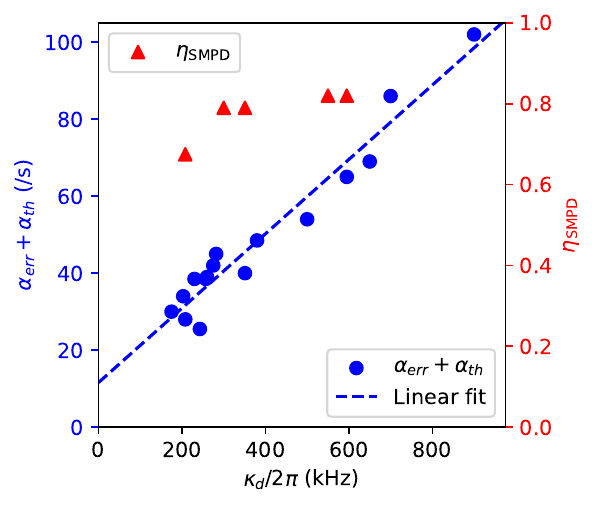}
\captionsetup{justification=Justified}
\caption{The blue full circles are the measured total dark counts, $\alpha_\mathrm{err} + \alpha_{\mathrm{th}}$, as a function of SMPD bandwidth $\kappa_d$. The dashed line is a linear fit, with slope $1.5\cdot 10^{-5}$, and intercept $11.6 \mathrm{s}^{-1}$. The red triangles show the measured efficiency at resonance, $\eta_\mathrm{SMPD}$.}
\label{fig:bdw_counts_eff}
\end{figure}

We finally study the SMPD figures of merit as a function of detection bandwidth, at the lowest temperature of $10$\,mK. We measure the maximum value of $\eta_\mathrm{SMPD}(\omega)$ and $\alpha$ as a function of $\kappa_d$ (see Fig.~\ref{fig:bdw_counts_eff}). We observe that $\eta_\mathrm{SMPD}$ increases with $\kappa_d$, although the dependence is very weak as soon as the SMPD bandwidth is above $250$\,kHz. This is expected because the measurements are performed with a monochromatic tone.
The reduction in efficiency at the lowest value of $\kappa_d$ may be attributed to the contribution of internal buffer losses, as well as possible transient effects that effectively shorten the detection time $T_d$. 

The false-positive rate $\alpha$, however, scales linearly with $\kappa_d$, as expected from Eq.~\ref{eq:th_countrate}. The fitted slope is $1.6 \cdot 10^{-5}$, in agreement with the expected $\eta_{\mathrm{SMPD}} \bar{n}_{\mathrm{th},b}/4$ yielding a thermal mode occupation $\bar{n}_{\mathrm{th},b} = 0.8 \cdot 10^{-4}$, slightly lower than in the measurements of Fig.~\ref{fig:BE} and corresponding to a mode temperature of $38$\,mK. The intercept at $\kappa_d = 0$ is $11.6\,\mathrm{s}^{-1}$, consistent with the error rate $\alpha_\mathrm{err}$ measured with pump detuned. Overall, the observed linear dependence of $\alpha_{\mathrm{th}}$ on $\kappa_d$ proves that these counts are indeed due to thermal photons impinging on the SMPD input line.
Not much can be inferred about their spectrum, apart from the fact that it appears to be flat on the scale of the detection bandwidth ($\sim 1$\,MHz).

For fig. \ref{fig:bdw_counts_eff}, no data have been taken above $\kappa_d/2\pi = $ 1 MHz. Indeed, above this value, high pump amplitudes are required to keep $C = 1$, and our microwave setup was unable to send high enough power.

\subsection{Single-spin measurements}

\begin{figure}[htpb!]
    \centering
    \includegraphics[width=0.98\linewidth]{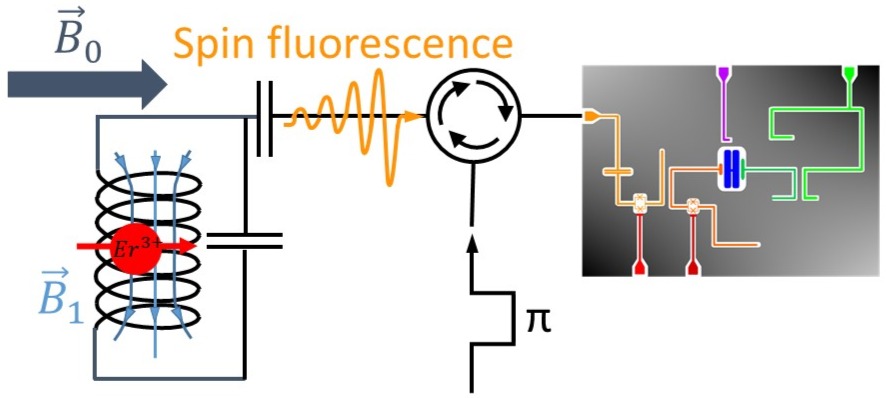}
    \includegraphics[width=0.98\linewidth]{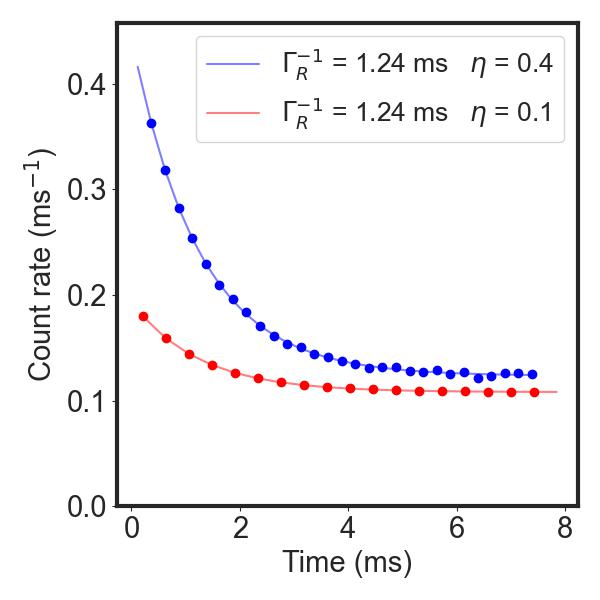}
    \captionsetup{justification=Justified}
    \caption{Microwave fluorescence of individual $\mathrm{Er}^{3+}$ electron spins in a $\mathrm{CaWO}_4$ host crystal matrix at $10$\,mK, detected by the SMPD reported here operated with $\kappa_d/2\pi = 500$ kHz (blue) and by a previous-generation SMPD operated with $\kappa_d/2\pi = 400$ kHz (red). Top : scheme of the experiment. The $Er^{3+}$ spin is magnetically coupled to the microwave field $\Vec{B}_1$ of an LC resonator and Zeeman tuned to resonance by a field $||\Vec{B}_0||$. The spin frequency is set to 7.7492 GHz in the latest version of the experiment, and 7.335 GHz for the "old" version. The spin is excited by a $\pi$-pulse, and the photon it emits upon radiative relaxation is detected by the SMPD. Graph : blue full circles are the measured count rate following a $\pi$ pulse excitation of the spin, as a function of the time, with the SMPD presented herein. The solid blue line is an exponential fit, from which we extract the spin lifetime $\Gamma_R^{-1} = 1.24$\,ms and spin-to-click efficiency, $\eta = 0.4$. Note that the noise floor $\alpha_{err}+\alpha_{th}=0.12 \mathrm{ms}^{-1}$ is slightly higher than expected from Fig.~\ref{fig:bdw_counts_eff}, possibly due to a different wiring scheme in the cryostat. The red full circles and solid line are the equivalent data and fit for the same experiment performed with a previous-generation SMPD. One can observe a fourfold improvement of $\eta$ from the previous to the newest SMPD generation, yielding a measurement time faster by an order of magnitude.}
    \label{fig:fluo_spin}
\end{figure}

We confirm the high efficiency of the detector by measuring the microwave fluorescence from a single electronic spin. The setup closely follows refs ~\cite{wang_single-electron_2023} and ~\cite{travesedo_all-microwave_2024}, and is schematically depicted in Fig.~\ref{fig:fluo_spin}. The electronic spin is an $\mathrm{Er}^{3+}$ ion in a $\mathrm{CaWO}_4$ matrix and is coupled to a superconducting resonator to enhance its radiative decay rate $\Gamma_R$. The spin is excited by a $\pi$ pulse with near-unit probability. It then relaxes radiatively by emitting a photon, which is either absorbed by the internal losses of the spin resonator, or emitted in the line, after which it may reach the SMPD detector or be absorbed by the losses between the two devices. The SMPD buffer $\omega_b$ is tuned to the spin and resonator frequency. The spin-to-click probability is therefore $\eta = \eta_{reso} \eta_{loss} \eta_\mathrm{SMPD}(\omega_b)$, with $\eta_{reso} = 0.6$ in this device (described in more detail elsewhere~\cite{travesedo_all-microwave_2024}). 

The spin microwave fluorescence curve is shown in Fig.~\ref{fig:fluo_spin}. This is well fitted by an exponential decay with time constant $\Gamma_R^{-1} = 1.24$ ms on top of a constant background, due to the SMPD dark counts. From the integral below the fluorescence curve, we obtain $\eta = 0.4$. This implies that the microwave power absorption between the spin device and the SMPD is $\eta_{loss} = 0.85$, a very plausible value given that the line includes a superconducting cable, a circulator, and an infrared filter. This provides an independent confirmation of the high efficiency reached by this SMPD device and of its relevance for single-spin spectroscopy.

\section{Conclusion}

We have demonstrated a microwave photon counter with tunable frequency (in the $7$\,GHz range) and bandwidth (between $\sim 100$\,kHz and $\sim 1$\,MHz). The device has a maximum efficiency of $0.8$, and its noise can be reduced by lowering the bandwidth. At an input bandwidth of $200$\,kHz, we demonstrate a record power sensitivity of $3 \cdot 10^{-23} \mathrm{W}/\sqrt{\mathrm{Hz}}$. The device is well suited to detecting emitters with a narrow line, and it will find applications in single-spin magnetic resonance~\cite{travesedo_all-microwave_2024,osullivan_nuclearspinregister2024} as well as in haloscope-based dark-matter searches~\cite{braggio_quantum-enhanced_2024}.

\section*{Acknowledgements}
{We acknowledge technical support from P.~S\'enat, D. Duet, P.-F.~Orfila, and S.~Delprat, and we are grateful for fruitful discussions within the Quantronics group. We acknowledge support from the Agence Nationale de la Recherche (ANR) through the MIRESPIN (ANR-19-CE47-0011) project. We acknowledge support of the R\'egion Ile-de-France through the DIM QUANTIP, from the AIDAS virtual joint laboratory, and from the France 2030 plan under Grant No. ANR-22-PETQ-0003. This project has received funding from the European Union Horizon 2020 research and innovation program under the OpenSuperQ100+ project, and from the European Research Council under Grant No. 101042315 (INGENIOUS). We acknowledge IARPA and Lincoln Labs for providing the Josephson traveling-wave parametric amplifier.}

\clearpage

\begin{figure}[htpb!]
\includegraphics[scale=0.575]{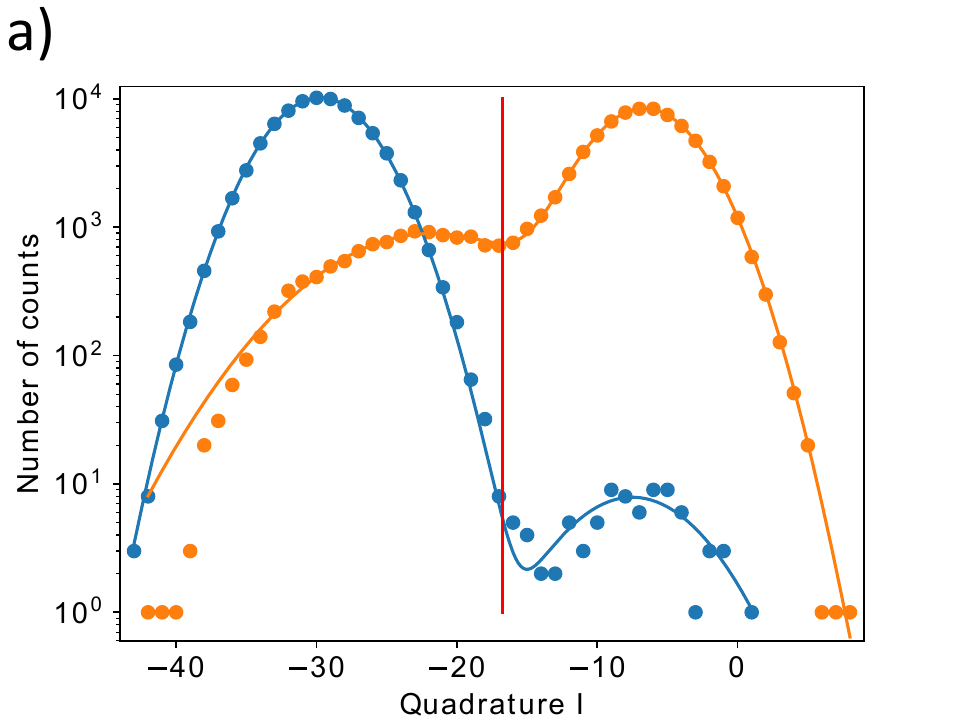}
\includegraphics[scale=0.55, trim = 0cm 0cm 0cm 0cm, clip]{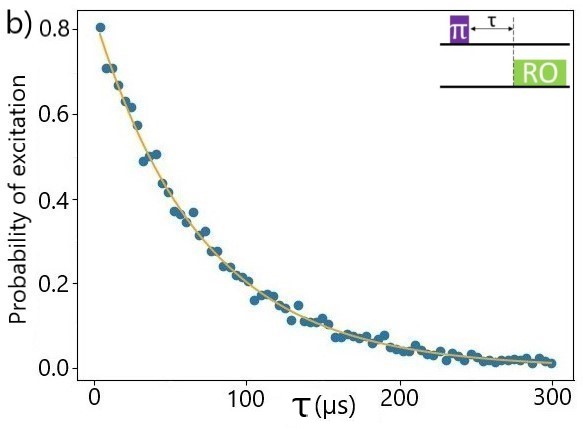}
\includegraphics[scale=0.51, trim = 0cm 0cm 0cm 0cm, clip]{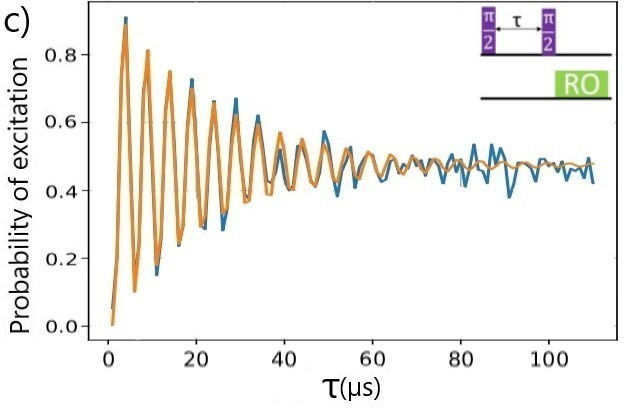}
\captionsetup{justification=Justified}
\caption{a) Qubit readout histograms after 95650 sequences, in the absence of any preparation pulse (blue full circles), and after a $\pi$ pulse (orange full circles). Solid lines are fits by a sum of two Gaussians centered on the two $I$ quadrature values corresponding to the ground and excited states, yielding $p_{th,q} = 8.5 \cdot 10^{-4}$, and $\mathcal{F}_{\mathrm{RO}}^{\ket{e}} = 0.87$. The red line shows the threshold. b) Qubit excited-state probability following a $\pi$ pulse, yielding $T_1 = 70$ µs. c) Qubit Ramsey fringes yielding $T_2^*=20$ µs. Insets : corresponding pulse sequences. Purple pulses are sent through the drive line, green through the readout (RO) line.}
\label{fig:qubit_parameters}
\end{figure}

\section*{Appendix A : Pump calibration}

\begin{figure*}[htpb!]
\includegraphics[scale=0.32]{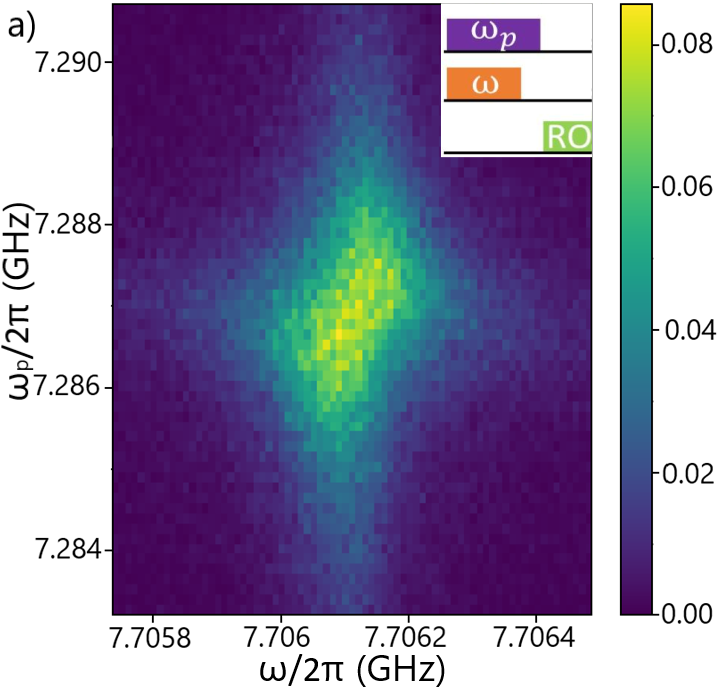}
\includegraphics[scale=0.31]{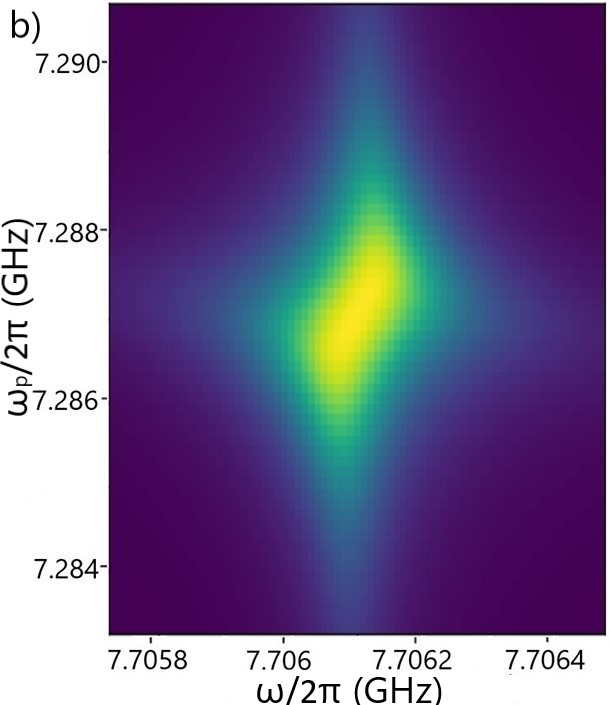}
\includegraphics[scale=0.5]{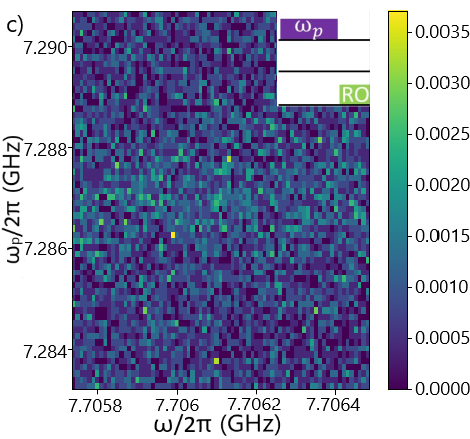}
\captionsetup{justification=Justified}
\caption{a) Four-wave mixing pattern : this color plot shows the measured qubit excited state as a function of pump frequency $\omega_p$ and signal frequency $\omega$ (from the input line). The qubit is excited only when the signal is close to $\omega_b$, and the pump verifies the 4WM condition. b) Fit from a two-coupled-cavity model. To fit the data, the value of the cooperativity $C$ is adjusted. Here, the value is $C=0.99$ (see appendix A). Other fit parameters are the buffer linewidth, $\nicefrac{\kappa_b}{2\pi} = 120$ kHz, and the waste linewidth, $\nicefrac{\kappa_w}{2\pi} = 1.75$ MHz. c) Same experiment as in a, but with the signal tone switched off. The higher excitation probabilities seen in subfigure a are no longer visible, demonstrating that we were indeed observing the actual 4WM pattern. Insets : corresponding pulse sequences. Purple pulses are sent through the pump/drive line, orange pulses through the signal/input line and green pulses through the waste/readout (RO) line. The pumping, signal, and readout pulses are respectively 15 µs, 10.5 µs and 800 ns long.}
\label{fig:4WM}
\end{figure*}

In this appendix we detail the tuning procedure for the pump parameters. The optimal pump frequency should satisfy the 4WM condition, $\omega_{\mathrm{4WM}} = \omega_{q}+\omega_{w}-\chi_{w}-\omega_{b}$. In the calibration procedure of the device, pump calibration comes after qubit characterization and readout calibration, which results are displayed in Fig. 6.

The pump amplitude is chosen by recording two-dimensional plots such as those shown in Fig. \ref{fig:4WM}, where the qubit excited-state probability is recorded as a function of pump and signal frequency, $\omega_p$ and $\omega$, respectively, for various pump powers. Each plot is then fitted to the following formula~\cite{albertinale_PhD_manuscript} :

\begin{equation}
|S_{w \leftarrow b}(\delta,\delta_{p})|^2 = \frac{4C}{\bigg | 1 + C - 4 \frac{\delta (\delta + \delta_{p})}{\kappa_{b} \kappa_{w}} + 2i \frac{\delta}{\kappa_{b}} + 2i \frac{\delta + \delta_{p}}{\kappa_{w}} \bigg |^2}
\label{eq:4WM_transmission}
\end{equation}

where $\delta = \omega - \omega_b$ is the signal-buffer detuning, and $\delta_{p} = \omega_p - \omega_{\mathrm{4WM}}$ the pump detuning from the 4WM condition.
\newline
If all detunings are zero, eq. \ref{eq:4WM_transmission} becomes :

\begin{equation}
|S_{w \leftarrow b}(\delta=0,\delta_{p}=0)|^2 = \frac{4C}{(1+C)^2}
\label{eq:4WM_transmission_tuned}
\end{equation}

The maximum of eq. \ref{eq:4WM_transmission_tuned} is reached for $C=1$ and its value is 1. \newline
The fit returns the value of $C$, and after a few iterations the condition $C=1$ is readily achieved. \newline
Note that the full width at half maximum along the signal frequency axis of $|S_{w \leftarrow b}(\delta,\delta_{p})|^2$ is the detection bandwidth of the SMPD:
\begin{equation}
\kappa_{d} = \sqrt{2}\sqrt{\sqrt{\kappa_{b}^2 \kappa_{w}^2 + \left ( \frac{\kappa_{b}-\kappa_{w}}{2} \right )^4}-\left ( \frac{\kappa_{b}-\kappa_{w}}{2} \right )^2}
\label{eq:Detection_bdw}
\end{equation}
When $\kappa_b \ll \kappa_w$ (as is the case in this work), the formula can be approximated as
\begin{equation}
\kappa_{d} \approx 2 \kappa_{b} 
\label{eq:Detection_smallbdw}
\end{equation}

\begin{figure}[htpb!]
    \centering
    \includegraphics[width=0.98\linewidth]{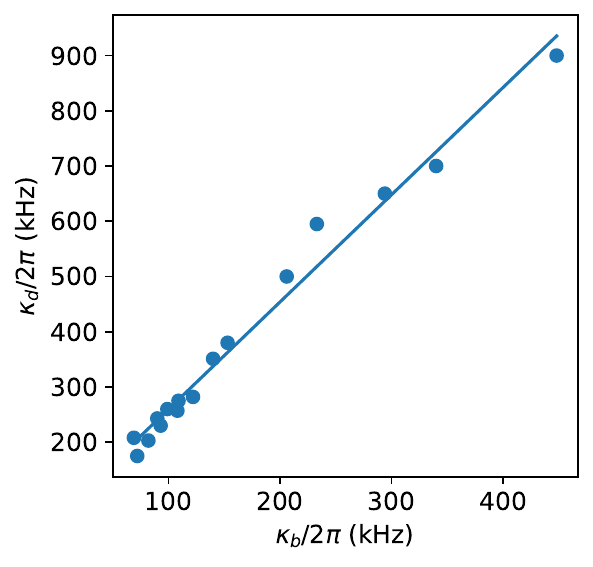}
    \captionsetup{justification=Justified}
    \caption{Detection bandwidth $\kappa_d$ as a function of buffer bandwidth $\kappa_b$. Full blue circles are experimental data and the blue solid line is a linear fit. The fit returns a slope of 1.9, which is close to 2, the value that is expected from eq. \ref{eq:Detection_smallbdw}}.
    \label{fig:k_d_vs_k_b}
\end{figure}

\noindent Fig. \ref{fig:k_d_vs_k_b} shows that eq. \ref{eq:Detection_smallbdw} is verified experimentally.

\section*{Appendix B : Efficiency measurement}

Measuring the efficiency of the SMPD begins with estimating the power at the buffer input. This is achieved by using the qubit itself as a detector. Ramsey fringes are shifted by $\Delta_q$ and dephased by $\Gamma_q$ according to the equation:

\begin{equation}
\Delta_{q} + i \Gamma_{q} = \frac{-4 \chi_{b} |\epsilon|^2}{(\kappa_{b}+i \chi_{b})^2 + 4 \delta^2},
\label{eq:Qubit_decoherence}
\end{equation}

\noindent where $\epsilon$ is the complex amplitude of the intracavity field. The efficiency is then simply computed by dividing the number of clicks per second by the number of photons per second at the device input.

\section*{Appendix C : SMPD fabrication}
We now describe the fabrication recipe of the SMPD presented in this paper.

\subsection*{C1 : Substrate preparation}
The process starts with a two-inch, double-side-polished sapphire wafer. This wafer is dipped in a piranha solution made with 2/3 sulfuric acid and 1/3 hydrogen peroxyde (by volume) and left for 20 minutes. Then, it is rinsed with water and dried with a stream of nitrogen. After that, the wafer is loaded into an RF reactive sputterer together with a tantalum target. The targeted thickness of the tantalum sputtering is between 80 and 90 nm. The sputtering process shall be performed while the wafer is heated up to 600°C to obtain alpha phase tantalum. \newline
The wafer is then prepared for dicing. To this end, UV III is spin-coated over the wafer for 60 seconds at 3000 rpm with an acceleration of 2000 rpm/s, followed by a postbake at 120°C for 3 minutes.
\subsection*{C2 : Circuit etching}
After dicing, the protective UV III layer is removed applying isopropanol (IPA) from a wash bottle the chosen chip for 15 seconds, then again for 5 seconds. For further cleaning, the chip is immersed in acetone under strong ultrasonic agitation for 5 minutes, then in another ultrasonic bath of IPA for 3 minutes. In the following, it is then dried with a stream of nitrogen and immersed in a piranha solution for 20 minutes. The chip is then rinsed with water, dehydrated with IPA, dried with a stream of nitrogen, and further dehydrated on a hotplate at 120°C for 5 minutes. Next, AZ1518 is spin-coated for 60 seconds at 3000 rpm with an acceleration of 2000 rpm/s. The postbake is performed on a hotplate for 2 minutes at 115°C. The circuit is then patterned using a UV LASER direct writer. A supplementary exposure is needed at the edge bead. The development is performed in pure MF319 for 60 seconds and stopped with water. The chip is dried with a stream of nitrogen and hard baked on a hotplate for 2 minutes at 115°C. Finally, the circuit is etched by dipping in Transene 111 (1/3 nitric acid, 1/3 hydrofluoric acid, and 1/3 water) for exactly 20 seconds. There is no need for agitation during etching. Rinsing is performed in a first bath of water for 20 seconds with agitation and then in a second bath of water for transport between benches.
\subsection*{C3 : Chip cleaning and oxide removal}
After the etching step, the chip is re-cleaned before proceeding to the e-beam related steps. Recently, we have added an oxide-removal step to the process which has not been applied to the sample presented here. First, the sample is immersed in a bath of acetone for 5 minutes to remove AZ1518, then in another bath of IPA for 3 minutes, and finally in piranha solution for 2 minutes. The chip is rinsed with water, dehydrated with IPA and dried with a stream of nitrogen. \newline
In the recently added oxidation step, the chip is immersed for 20 minutes in buffered oxide etchant (BOE), rinsed in a first bath of water for 20 seconds with agitation, then in a second bath of water for transport between benches, dehydrated with IPA, dried with a stream of nitrogen, and further dehydrated on a hotplate during 5 minutes at 120°C.
\subsection*{C4 : Preparation of e-beam bilayer}
A first layer of MAA EL-6 is spin-coated for 60 seconds at 4000 rpm with an acceleration of 1000 rpm/s. The postbake is performed on a hotplate for 5 minutes at 172°C. This is followed by spin coating of a second layer of PMMA 950k A6 for 60 seconds at 3000 rpm with an acceleration of 2000 rpm/s. The postbake, on a hotplate at 175°C, lasts 15 minutes. \newline
The substrate being made of sapphire, there is no way for electrons to escape through it during lithography. Therefore, the bilayer must be metallized to prevent charge effects. For this reason, a 7-nm thick layer of aluminum is evaporated over the bilayer at a rate of 0.2 nm/s. There is no need for a good vacuum for this evaporation step ; a pressure of about $10^{-6}$ hPa is enough.
\subsection*{C5 : E-beam lithography}
Given the need for the simultaneous presence of Josephson junctions of very different sizes across the device, the Manhattan process has been chosen. Due to the absence of a 100-kV electron beam lithography (EBL) system in our laboratory, we developed an adapted version of this process using a bilayer with a reversed aspect ratio (a layer of MAA about 170 nm thick beneath a layer of PMMA about 580 nm thick). We use the following parameters with our Raith e-LINE system :
\begin{itemize}
    \item Magnification : $\times 1000$.
    \item Base dose : $250 \mu C/cm^2$.
    \item Relative dose factor : 1.5 for "big" junctions' main pattern (about 1 µm edge size), 1.7 for "small" junctions' main pattern (about 200 nm edge size or less), 0.3 for undercut boxes.
    \item Voltage : 25 kV.
    \item Aperture : 10 µm.
    \item Current : About 20 pA.
    \item Writefield : $100 \times 100 \mu m^2$.
\end{itemize}
\subsection*{C6 : Aluminum etching, resist development, and descum}
Before developing the e-beam resist, the metallization layer must be removed. Aluminum etching is performed by dipping the chip in a 10 g/l potassium hydroxyde (KOH) solution in water for 30 seconds. It is then rinsed by for 3 seconds in water, followed by 3 seconds in IPA. The development itself begins immediately after : The chip is immersed for 45 seconds in a solution made of 1/4 methyl isobutyl ketone (MIBK) and 3/4 IPA (by volume). Development is stopped by dipping in IPA for 15 seconds. The chip and the bilayer are dried first with a stream of nitrogen, then on a hotplate at 70°C for 1 minute. Finally, the resist mask is cleaned in an oxygen plasma asher for 10 seconds, at a pressure of 0.2 hPa and an ionizing power of 75 W.
\subsection*{C7 : Josephson junctions evaporation}
For evaporating Josephson junctions, a good vacuum is required (at least below $5 \cdot 10^{-7}$ hPa). Given that we are using the Manhattan process, the evaporator must be equipped with a planetary rotation. The parameters given here apply to one of our Plassys MEB 550S systems. \newline
First, the surface is prepared by argon-ion milling with an incidence angle of 0°. The ion gun is tuned with the filament at 500 V and 25 mA, and the acceleration grid at 100 V. The shutter is removed from the ion-beam trajectory twice for 5 seconds each, with 1 minute of rest between exposures. Then comes the evaporation of the first layer of aluminum, performed at an angle of 30° and a rate of 1 nm/s. A thickness of 29 nm is set on the evaporator to obtain an actual thickness of 25 nm. This first layer is oxidized by a mixture of oxygen and argon at 10 hPa for 5 minutes. The sample holder is rotated around the planetary axis by 90° in order to evaporate along the orthogonal direction of the Manhattan pattern. Finally, the second evaporation of aluminum is performed with the same parameters (30° angle and 1 nm/s rate), except that a thickness of 69 nm is is set to get an actual thickness of 60 nm.
\subsection*{C8 : Final lift off}
The lift off is performed in two successive baths of acetone at 60°C. Blowing with a Pasteur pipette helps for removal. The sample is then brought back to room temperature with a non-heated bath of acetone. It is then dehydrated in IPA and finally dried by dinitrogen blow.

\section*{Appendix D : wiring diagram}

\begin{figure*}[htpb!]
    \centering
    \includegraphics[scale = 0.9]{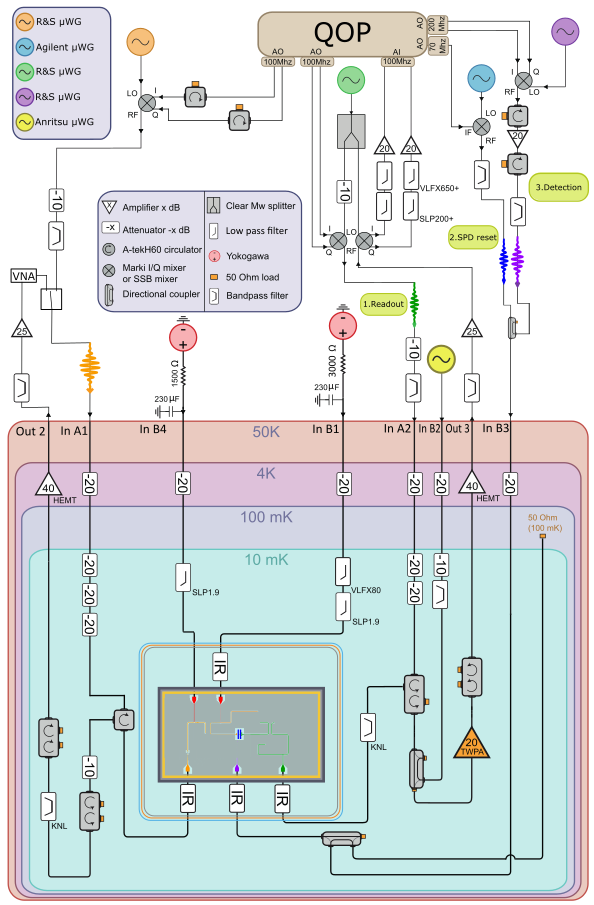}
    \captionsetup{justification=Justified}
    \caption{Wiring used for the characterization run reported herein.}
    \label{fig:wiring}
\end{figure*}

The wiring scheme used for SMPD characterization is presented in fig. \ref{fig:wiring}. Lines "In A1" and "Out 2" are used to probe the buffer resonator in reflection (to determine $\omega_b$ and $\kappa_b$), and also to calibrate the detector efficiency by sending pulses containing a well-controlled number of photons. "In B1" and "In B4" are flux bias lines for controlling the SQUID inductances of the buffer and its Purcell resonator, respectively. "In A2" and "Out 3" are used for probing the waste resonator in reflection (to determine $\omega_w$ and $\kappa_w$), and also to perform dispersive readout of the qubit through the waste resonator. "In B2" is used to pump the travelling-wave parametric amplifier (TWPA) on the "Out 3" line. "In B3" is the qubit drive line.

\clearpage

\bibliography{bib2}

@misc{travesedo_all-microwave_2024,
	title = {All-microwave readout, spectroscopy, and dynamic polarization of individual nuclear spins in a crystal},
	url = {http://arxiv.org/abs/2408.14282},
	doi = {10.48550/arXiv.2408.14282},
	urldate = {2024-09-15},
	publisher = {arXiv},
	author = {Travesedo, J. and O'Sullivan, J. and Pallegoix, L. and Huang, Z. W. and Hogan, P. and Goldner, P. and Chaneliere, T. and Bertaina, S. and Esteve, D. and Abgrall, P. and Vion, D. and Flurin, E. and Bertet, P.},
	month = aug,
	year = {2024},
	note = {arXiv:2408.14282 [cond-mat, physics:quant-ph]},
}

@misc{braggio_quantum-enhanced_2024,
	title = {Quantum-enhanced sensing of axion dark matter with a transmon-based single microwave photon counter},
	url = {http://arxiv.org/abs/2403.02321},
	doi = {10.48550/arXiv.2403.02321},
	urldate = {2024-09-15},
	publisher = {arXiv},
	author = {Braggio, C. and Balembois, L. and Di Vora, R. and Wang, Z. and Travesedo, J. and Pallegoix, L. and Carugno, G. and Ortolan, A. and Ruoso, G. and Gambardella, U. and D'Agostino, D. and Bertet, P. and Flurin, E.},
	month = mar,
	year = {2024},
	note = {arXiv:2403.02321 [hep-ex, physics:quant-ph]},
}

@misc{osullivan_nuclearspinregister2024,
	title = {Individual solid-state nuclear spin qubits with coherence exceeding seconds},
	author = {O'Sullivan, J. and Travesedo, J. and Pallegoix, L. and Huang, Z.W. and Hogan, P. and Goldner, P. and Esteve, D. and Vion, D. and Bertet, P. and Flurin, E.},
    publisher = {arXiv},
    url = {https://arxiv.org/pdf/2410.10432},
	doi = {https://doi.org/10.48550/arXiv.2410.10432},
	year = {2024},
}

@article{dixit_searching_2021,
	title = {Searching for {Dark} {Matter} with a {Superconducting} {Qubit}},
	volume = {126},
	url = {https://link.aps.org/doi/10.1103/PhysRevLett.126.141302},
	doi = {10.1103/PhysRevLett.126.141302},
	number = {14},
	urldate = {2024-10-08},
	journal = {Physical Review Letters},
	author = {Dixit, Akash V. and Chakram, Srivatsan and He, Kevin and Agrawal, Ankur and Naik, Ravi K. and Schuster, David I. and Chou, Aaron},
	month = apr,
	year = {2021},
	note = {Publisher: American Physical Society},
	pages = {141302},
}

@article{lee_graphene-based_2020,
	title = {Graphene-based {Josephson} junction microwave bolometer},
	volume = {586},
	copyright = {2020 The Author(s), under exclusive licence to Springer Nature Limited},
	issn = {1476-4687},
	url = {https://www.nature.com/articles/s41586-020-2752-4},
	doi = {10.1038/s41586-020-2752-4},
	language = {english},
	number = {7827},
	urldate = {2024-02-25},
	journal = {Nature},
	author = {Lee, Gil-Ho and Efetov, Dmitri K. and Jung, Woochan and Ranzani, Leonardo and Walsh, Evan D. and Ohki, Thomas A. and Taniguchi, Takashi and Watanabe, Kenji and Kim, Philip and Englund, Dirk and Fong, Kin Chung},
	month = oct,
	year = {2020},
	note = {Number: 7827
Publisher: Nature Publishing Group},
	pages = {42--46},
}

@article{royer_itinerant_2018,
	title = {Itinerant {Microwave} {Photon} {Detector}},
	volume = {120},
	url = {https://link.aps.org/doi/10.1103/PhysRevLett.120.203602},
	doi = {10.1103/PhysRevLett.120.203602},
	number = {20},
	urldate = {2024-10-04},
	journal = {Physical Review Letters},
	author = {Royer, Baptiste and Grimsmo, Arne L. and Choquette-Poitevin, Alexandre and Blais, Alexandre},
	month = may,
	year = {2018},
	note = {Publisher: American Physical Society},
	pages = {203602},
}

@article{gu_microwave_2017,
	series = {Microwave photonics with superconducting quantum circuits},
	title = {Microwave photonics with superconducting quantum circuits},
	volume = {718-719},
	issn = {0370-1573},
	url = {https://www.sciencedirect.com/science/article/pii/S0370157317303290},
	doi = {10.1016/j.physrep.2017.10.002},
	urldate = {2024-10-04},
	journal = {Physics Reports},
	author = {Gu, Xiu and Kockum, Anton Frisk and Miranowicz, Adam and Liu, Yu-xi and Nori, Franco},
	month = nov,
	year = {2017},
	pages = {1--102},
}

@article{sathyamoorthy_detecting_2016,
	title = {Detecting itinerant single microwave photons},
	volume = {17},
	issn = {1878-1535},
	url = {https://comptes-rendus.academie-sciences.fr/physique/articles/10.1016/j.crhy.2016.07.010/},
	doi = {10.1016/j.crhy.2016.07.010},
	language = {french},
	number = {7},
	urldate = {2024-10-04},
	journal = {Comptes Rendus. Physique},
	author = {Sathyamoorthy, Sankar Raman and Stace, Thomas M. and Johansson, Göran},
	year = {2016},
	pages = {756--765},
}

@article{kyriienko_continuous-wave_2016,
	title = {Continuous-{Wave} {Single}-{Photon} {Transistor} {Based} on a {Superconducting} {Circuit}},
	volume = {117},
	url = {https://link.aps.org/doi/10.1103/PhysRevLett.117.140503},
	doi = {10.1103/PhysRevLett.117.140503},
	number = {14},
	urldate = {2024-10-04},
	journal = {Physical Review Letters},
	author = {Kyriienko, Oleksandr and Sørensen, Anders S.},
	month = sep,
	year = {2016},
	note = {Publisher: American Physical Society},
	pages = {140503},
}

@article{sathyamoorthy_quantum_2014,
	title = {Quantum {Nondemolition} {Detection} of a {Propagating} {Microwave} {Photon}},
	volume = {112},
	url = {https://link.aps.org/doi/10.1103/PhysRevLett.112.093601},
	doi = {10.1103/PhysRevLett.112.093601},
	number = {9},
	urldate = {2024-10-04},
	journal = {Physical Review Letters},
	author = {Sathyamoorthy, Sankar R. and Tornberg, L. and Kockum, Anton F. and Baragiola, Ben Q. and Combes, Joshua and Wilson, C. M. and Stace, Thomas M. and Johansson, G.},
	month = mar,
	year = {2014},
	note = {Publisher: American Physical Society},
	pages = {093601},
}

@article{romero_microwave_2009,
	title = {Microwave {Photon} {Detector} in {Circuit} {QED}},
	volume = {102},
	url = {https://link.aps.org/doi/10.1103/PhysRevLett.102.173602},
	doi = {10.1103/PhysRevLett.102.173602},
	number = {17},
	urldate = {2024-10-04},
	journal = {Physical Review Letters},
	author = {Romero, G. and García-Ripoll, J. J. and Solano, E.},
	month = apr,
	year = {2009},
	note = {Publisher: American Physical Society},
	pages = {173602},
}

@article{helmer_quantum_2009,
	title = {Quantum nondemolition photon detection in circuit {QED} and the quantum {Zeno} effect},
	volume = {79},
	url = {https://link.aps.org/doi/10.1103/PhysRevA.79.052115},
	doi = {10.1103/PhysRevA.79.052115},
	number = {5},
	urldate = {2024-09-28},
	journal = {Physical Review A},
	author = {Helmer, Ferdinand and Mariantoni, Matteo and Solano, Enrique and Marquardt, Florian},
	month = may,
	year = {2009},
	note = {Publisher: American Physical Society},
	pages = {052115},
}

@article{opremcak_measurement_2018,
	title = {Measurement of a superconducting qubit with a microwave photon counter},
	volume = {361},
	issn = {0036-8075, 1095-9203},
	url = {https://www.science.org/doi/10.1126/science.aat4625},
	doi = {10.1126/science.aat4625},
	language = {english},
	number = {6408},
	urldate = {2024-10-04},
	journal = {Science},
	author = {Opremcak, A. and Pechenezhskiy, I. V. and Howington, C. and Christensen, B. G. and Beck, M. A. and Leonard, E. and Suttle, J. and Wilen, C. and Nesterov, K. N. and Ribeill, G. J. and Thorbeck, T. and Schlenker, F. and Vavilov, M. G. and Plourde, B. L. T. and McDermott, R.},
	month = sep,
	year = {2018},
	pages = {1239--1242},
}

@article{besse_single-shot_2018,
	title = {Single-{Shot} {Quantum} {Nondemolition} {Detection} of {Individual} {Itinerant} {Microwave} {Photons}},
	volume = {8},
	url = {https://link.aps.org/doi/10.1103/PhysRevX.8.021003},
	doi = {10.1103/PhysRevX.8.021003},
	number = {2},
	urldate = {2024-10-04},
	journal = {Physical Review X},
	author = {Besse, Jean-Claude and Gasparinetti, Simone and Collodo, Michele C. and Walter, Theo and Kurpiers, Philipp and Pechal, Marek and Eichler, Christopher and Wallraff, Andreas},
	month = apr,
	year = {2018},
	note = {Publisher: American Physical Society},
	pages = {021003},
}

@article{briegel_measurement-based_2009,
	title = {Measurement-based quantum computation},
	volume = {5},
	copyright = {2009 Springer Nature Limited},
	issn = {1745-2481},
	url = {https://www.nature.com/articles/nphys1157},
	doi = {10.1038/nphys1157},
	language = {english},
	number = {1},
	urldate = {2024-10-04},
	journal = {Nature Physics},
	author = {Briegel, H. J. and Browne, D. E. and Dür, W. and Raussendorf, R. and Van den Nest, M.},
	month = jan,
	year = {2009},
	note = {Publisher: Nature Publishing Group},
	pages = {19--26},
}

@misc{bartolucci_fusion-based_2021,
	title = {Fusion-based quantum computation},
	url = {https://arxiv.org/abs/2101.09310v1},
	language = {english},
	urldate = {2024-10-04},
	journal = {arXiv.org},
	author = {Bartolucci, Sara and Birchall, Patrick and Bombin, Hector and Cable, Hugo and Dawson, Chris and Gimeno-Segovia, Mercedes and Johnston, Eric and Kieling, Konrad and Nickerson, Naomi and Pant, Mihir and Pastawski, Fernando and Rudolph, Terry and Sparrow, Chris},
	month = jan,
	year = {2021},
}

@article{raussendorf_measurement-based_2003,
	title = {Measurement-based quantum computation on cluster states},
	volume = {68},
	url = {https://link.aps.org/doi/10.1103/PhysRevA.68.022312},
	doi = {10.1103/PhysRevA.68.022312},
	number = {2},
	urldate = {2024-10-04},
	journal = {Physical Review A},
	author = {Raussendorf, Robert and Browne, Daniel E. and Briegel, Hans J.},
	month = aug,
	year = {2003},
	note = {Publisher: American Physical Society},
	pages = {022312},
}

@article{narla_robust_2016,
	title = {Robust {Concurrent} {Remote} {Entanglement} {Between} {Two} {Superconducting} {Qubits}},
	volume = {6},
	url = {https://link.aps.org/doi/10.1103/PhysRevX.6.031036},
	doi = {10.1103/PhysRevX.6.031036},
	number = {3},
	urldate = {2024-10-04},
	journal = {Physical Review X},
	author = {Narla, A. and Shankar, S. and Hatridge, M. and Leghtas, Z. and Sliwa, K. M. and Zalys-Geller, E. and Mundhada, S. O. and Pfaff, W. and Frunzio, L. and Schoelkopf, R. J. and Devoret, M. H.},
	month = sep,
	year = {2016},
	note = {Publisher: American Physical Society},
	pages = {031036},
}

@article{assouly_quantum_2023,
	title = {Quantum advantage in microwave quantum radar},
	volume = {19},
	copyright = {2023 The Author(s), under exclusive licence to Springer Nature Limited},
	issn = {1745-2481},
	url = {https://www.nature.com/articles/s41567-023-02113-4},
	doi = {10.1038/s41567-023-02113-4},
	language = {english},
	number = {10},
	urldate = {2024-10-04},
	journal = {Nature Physics},
	author = {Assouly, R. and Dassonneville, R. and Peronnin, T. and Bienfait, A. and Huard, B.},
	month = oct,
	year = {2023},
	note = {Publisher: Nature Publishing Group},
	pages = {1418--1422},
}

@article{scigliuzzo_primary_2020,
	title = {Primary {Thermometry} of {Propagating} {Microwaves} in the {Quantum} {Regime}},
	volume = {10},
	url = {https://link.aps.org/doi/10.1103/PhysRevX.10.041054},
	doi = {10.1103/PhysRevX.10.041054},
	number = {4},
	urldate = {2024-10-04},
	journal = {Physical Review X},
	author = {Scigliuzzo, Marco and Bengtsson, Andreas and Besse, Jean-Claude and Wallraff, Andreas and Delsing, Per and Gasparinetti, Simone},
	month = dec,
	year = {2020},
	note = {Publisher: American Physical Society},
	pages = {041054},
}

@article{lamoreaux_analysis_2013,
	title = {Analysis of single-photon and linear amplifier detectors for microwave cavity dark matter axion searches},
	volume = {88},
	url = {https://link.aps.org/doi/10.1103/PhysRevD.88.035020},
	doi = {10.1103/PhysRevD.88.035020},
	number = {3},
	urldate = {2024-10-04},
	journal = {Physical Review D},
	author = {Lamoreaux, S. K. and van Bibber, K. A. and Lehnert, K. W. and Carosi, G.},
	month = aug,
	year = {2013},
	note = {Publisher: American Physical Society},
	pages = {035020},
}

@article{gisin_quantum_2002,
    title = {Quantum cryptography},
    volume = {74},
    url = {https://journals.aps.org/rmp/pdf/10.1103/RevModPhys.74.145},
    doi = {https://doi.org/10.1103/RevModPhys.74.145},
    journal = {Review of Modern Physics},
    author = {Gisin, N. and Ribordy, G. and Tittel, W. and Zbinden,  H.},
    month = jan,
    year = {2002},
    note = {Publisher : American Physical Society},
}

@article{besse_parity_2020,
    author = {J.-C. Besse and S. Gasparinetti and M. C. Collodo and T. Walter and A. Remm and J. Krause and C. Eichler and A. Wallraff},
    title = {Parity Detection of Propagating Microwave Fields},
    journal = {Physical Review X},
    year = {2020},
}

@phdthesis{albertinale_PhD_manuscript,
    author = {Emanuele Albertinale},
    title = {Measuring spin fluorescence with a microwave photon detector},
    school = {Université Paris-Saclay},
    year = {2021},
}

@article{blais_cqed_2021,
    author = {Blais, A. and Grimsmo, A. L. and Girvin, S.M. and Wallraff, A.},
    title = {Circuit quantum electrodynamics},
    journal = {Reviews of Modern Physics},
    year = {2021}
}

@article{orrit_single_1990,
	title = {Single pentacene molecules detected by fluorescence excitation in a p-terphenyl crystal},
	volume = {65},
	url = {https://link.aps.org/doi/10.1103/PhysRevLett.65.2716},
	doi = {10.1103/PhysRevLett.65.2716},
	number = {21},
	urldate = {2024-10-04},
	journal = {Physical Review Letters},
	author = {Orrit, M. and Bernard, J.},
	month = nov,
	year = {1990},
	note = {Publisher: American Physical Society},
	pages = {2716--2719},
}

@article{klar_fluorescence_2000,
	title = {Fluorescence microscopy with diffraction resolution barrier broken by stimulated emission},
	volume = {97},
	issn = {0027-8424},
	doi = {10.1073/pnas.97.15.8206},
	language = {english},
	number = {15},
	journal = {Proceedings of the National Academy of Sciences of the United States of America},
	author = {Klar, T. A. and Jakobs, S. and Dyba, M. and Egner, A. and Hell, S. W.},
	month = jul,
	year = {2000},
	pmid = {10899992},
	pmcid = {PMC26924},
	pages = {8206--8210},
}

@article{betzig_imaging_2006,
	title = {Imaging intracellular fluorescent proteins at nanometer resolution},
	volume = {313},
	issn = {1095-9203},
	doi = {10.1126/science.1127344},
	language = {english},
	number = {5793},
	journal = {Science (New York, N.Y.)},
	author = {Betzig, Eric and Patterson, George H. and Sougrat, Rachid and Lindwasser, O. Wolf and Olenych, Scott and Bonifacino, Juan S. and Davidson, Michael W. and Lippincott-Schwartz, Jennifer and Hess, Harald F.},
	month = sep,
	year = {2006},
	pmid = {16902090},
	pages = {1642--1645},
}

@article{bruschini_single-photon_2019,
	title = {Single-photon avalanche diode imagers in biophotonics: review and outlook},
	volume = {8},
	issn = {2047-7538},
	shorttitle = {Single-photon avalanche diode imagers in biophotonics},
	doi = {10.1038/s41377-019-0191-5},
	language = {english},
	journal = {Light, Science \& Applications},
	author = {Bruschini, Claudio and Homulle, Harald and Antolovic, Ivan Michel and Burri, Samuel and Charbon, Edoardo},
	year = {2019},
	pmid = {31645931},
	pmcid = {PMC6804596},
	pages = {87},
}

@article{lescanne_irreversible_2020,
	title = {Irreversible {Qubit}-{Photon} {Coupling} for the {Detection} of {Itinerant} {Microwave} {Photons}},
	volume = {10},
	issn = {2160-3308},
	url = {https://link.aps.org/doi/10.1103/PhysRevX.10.021038},
	doi = {10.1103/PhysRevX.10.021038},
	language = {english},
	number = {2},
	urldate = {2023-02-23},
	journal = {Physical Review X},
	author = {Lescanne, Raphaël and Deléglise, Samuel and Albertinale, Emanuele and Réglade, Ulysse and Capelle, Thibault and Ivanov, Edouard and Jacqmin, Thibaut and Leghtas, Zaki and Flurin, Emmanuel},
	month = may,
	year = {2020},
	pages = {021038},
}

@article{wang_single-electron_2023,
	title = {Single-electron spin resonance detection by microwave photon counting},
	volume = {619},
	copyright = {2023 The Author(s), under exclusive licence to Springer Nature Limited},
	issn = {1476-4687},
	url = {https://www.nature.com/articles/s41586-023-06097-2},
	doi = {10.1038/s41586-023-06097-2},
	language = {english},
	number = {7969},
	urldate = {2024-02-15},
	journal = {Nature},
	author = {Wang, Z. and Balembois, L. and Rančić, M. and Billaud, E. and Le Dantec, M. and Ferrier, A. and Goldner, P. and Bertaina, S. and Chanelière, T. and Esteve, D. and Vion, D. and Bertet, P. and Flurin, E.},
	month = jul,
	year = {2023},
	note = {Number: 7969
Publisher: Nature Publishing Group},
	pages = {276--281},
}

@article{balembois_cyclically2024,
  title = {Cyclically Operated Microwave Single-Photon Counter with Sensitivity of ${10}^{\ensuremath{-}22}\phantom{\rule{0.2em}{0ex}}\mathrm{W}/\sqrt{\mathrm{Hz}}$},
  author = {Balembois, L. and Travesedo, J. and Pallegoix, L. and May, A. and Billaud, E. and Villiers, M. and Est\`eve, D. and Vion, D. and Bertet, P. and Flurin, E.},
  journal = {Phys. Rev. Appl.},
  volume = {21},
  issue = {1},
  pages = {014043},
  numpages = {12},
  year = {2024},
  month = {Jan},
  publisher = {American Physical Society},
  doi = {10.1103/PhysRevApplied.21.014043},
  url = {https://link.aps.org/doi/10.1103/PhysRevApplied.21.014043}
}

@article{albertinale_detecting_2021,
	title = {Detecting spins by their fluorescence with a microwave photon counter},
	volume = {600},
	copyright = {2021 The Author(s), under exclusive licence to Springer Nature Limited},
	issn = {1476-4687},
	url = {https://www.nature.com/articles/s41586-021-04076-z},
	doi = {10.1038/s41586-021-04076-z},
	language = {english},
	number = {7889},
	urldate = {2024-09-15},
	journal = {Nature},
	author = {Albertinale, Emanuele and Balembois, Léo and Billaud, Eric and Ranjan, Vishal and Flanigan, Daniel and Schenkel, Thomas and Estève, Daniel and Vion, Denis and Bertet, Patrice and Flurin, Emmanuel},
	month = dec,
	year = {2021},
	note = {Publisher: Nature Publishing Group},
	pages = {434--438},
}

@article{kono_quantum_2018,
	title = {Quantum non-demolition detection of an itinerant microwave photon},
	volume = {14},
	issn = {1745-2473, 1745-2481},
	url = {http://arxiv.org/abs/1711.05479},
	doi = {10.1038/s41567-018-0066-3},
	number = {6},
	urldate = {2024-02-25},
	journal = {Nature Physics},
	author = {Kono, S. and Koshino, K. and Tabuchi, Y. and Noguchi, A. and Nakamura, Y.},
	month = jun,
	year = {2018},
	note = {arXiv:1711.05479 [quant-ph]},
	pages = {546--549},
}

@article{inomata_single_2016,
	title = {Single microwave-photon detector using an artificial \${\textbackslash}{Lambda}\$-type three-level system},
	volume = {7},
	issn = {2041-1723},
	url = {http://arxiv.org/abs/1601.05513},
	doi = {10.1038/ncomms12303},
	number = {1},
	urldate = {2024-02-25},
	journal = {Nature Communications},
	author = {Inomata, Kunihiro and Lin, Zhirong and Koshino, Kazuki and Oliver, William D. and Tsai, Jaw-Shen and Yamamoto, Tsuyoshi and Nakamura, Yasunobu},
	month = jul,
	year = {2016},
	note = {arXiv:1601.05513 [cond-mat, physics:quant-ph]},
	pages = {12303},
}

@article{billaud_fluorescence-detection_2023,
    author = {E. Billaud and L. Balembois and M. Le Dantec and M. Rančić and E. Albertinale and S. Bertaina and T. Chanelière and P. Goldner and D. Estève and D. Vion and P. Bertet and E. Flurin},
    title = {Microwave Fluorescence Detection of Spin Echoes},
    journal = {Physical Review Letters},
    year = {2023},
    url = {https://journals.aps.org/prl/pdf/10.1103/PhysRevLett.131.100804},
    doi = {https://doi.org/10.1103/PhysRevLett.131.100804}
}

@article{wong_double-quantum-dot_2017,
    author = {C. H. Wong and M. G. Vavilov},
    title = {Quantum efficiency of a single microwave photon detector based on a semiconductor double quantum dot},
    journal = {Physical Review A},
    year = {2017},
    url = {https://journals.aps.org/pra/pdf/10.1103/PhysRevA.95.012325},
    doi = {10.1103/PhysRevA.95.012325}
}

@article{chen_josephson-junctions-based_2011,
    author = {Y.-F. Chen and D. Hover and S. Sendelbach and L. Maurer and S. T. Merkel and E. J. Pritchett and F. K. Wilhelm and R. McDermott},
    title = {Microwave Photon Counter Based on Josephson Junctions},
    journal = {Physical Review Letters},
    year = {2011},
    url = {https://journals.aps.org/prl/pdf/10.1103/PhysRevLett.107.217401},
    doi = {10.1103/PhysRevLett.107.217401}
}

@article{koshino_impedance-matched_2013,
    author = {K. Koshino and K. Inomata and T. Yamamoto and Y. Nakamura},
    title = {Implementation of an Impedance-Matched $\Lambda$ System by Dressed-State Engineering},
    journal = {Physical Review Letter},
    year = {2013},
    url = {https://journals.aps.org/prl/pdf/10.1103/PhysRevLett.111.153601},
    doi = {10.1103/PhysRevLett.111.153601}
}

@article{Ghirri_microwave_2020,
    author = {A. Ghirri and S. Cornia and M. Affronte},
    title = {Microwave Photon Detectors Based on Semiconducting Double Quantum Dots},
    journal = {Sensors},
    year = {2020},
    url = {https://www.mdpi.com/1424-8220/20/14/4010},
    doi = {https://doi.org/10.3390/s20144010}
}

@article{walter_realizing_2017,
    author = {T. Walter and P. Kurpiers and S. Gasparinetti and P. Magnard and A. Potočnik and Y. Salathé and M. Pechal and M. Mondal and M. Oppliger and C. Eichler and A. Wallraff},
    title = {Realizing Rapid, High-Fidelity, Single-Shot Dispersive Readout of Superconducting Qubits},
    url = {https://journals.aps.org/prapplied/pdf/10.1103/PhysRevApplied.7.054020},
    doi = {https://doi.org/10.1103/PhysRevApplied.7.054020},
    journal = {Physical Review Applied},
    year = {2017}
}

@article{banchi_molecular_2020,
    author = {L. Banchi and M. Fingerhuth and T. Babej and C. Ing and J. M. Arrazola},
    title = {Molecular docking with Gaussian Boson Sampling},
    url = {https://www.science.org/doi/epdf/10.1126/sciadv.aax1950},
    journal = {Science Advances},
    year = {2020}
}

\end{document}